\documentclass[aps,twocolumn,floatfix,superscriptaddress,longbibliography,nofootinbib,showpacs]{revtex4}

\usepackage{amssymb,amsfonts,amsmath}
\usepackage{graphicx}
\usepackage{indentfirst}
\usepackage{latexsym}
\usepackage{tabls}
\usepackage{epsfig}
\usepackage{multirow}
\usepackage{subfigure}
\usepackage{comment}
\usepackage{hyperref}
\usepackage{color}
\usepackage[usenames,dvipsnames]{xcolor}
\usepackage{bm}
\usepackage{mathrsfs}
\usepackage{enumitem}
\usepackage[normalem]{ulem}

\newcommand{\Msun}{\ensuremath{M_{ \odot }}}
\newcommand{\simga}{\ensuremath{\sigma}}
\newcommand{\dd}{\ensuremath{\mathrm{d}}}

\DeclareMathOperator{\inv}{inv}

\DeclareMathOperator{\Ex}{\mathbb{E}}

\begin{document}

\title{Improving gravitational-wave parameter estimation using Gaussian process regression}

\def\addCambridge{Institute of Astronomy, Madingley Road, Cambridge, CB3 0HA, UK}
\def\addBirmingham{School of Physics \& Astronomy, University of Birmingham, Birmingham, B15 2TT, UK}
\def\addEdinburgh{School of Mathematics, University of Edinburgh, King's Buildings, Edinburgh, EH9 3JZ, UK}

\author{Christopher J.\ Moore}
\email{cjm96@ast.cam.ac.uk}
\affiliation{\addCambridge} 

\author{Christopher P.\ L.\ Berry}
\email{cplb@star.sr.bham.ac.uk}
\affiliation{\addBirmingham} 

\author{Alvin J.\ K.\ Chua}
\email{ajkc3@ast.cam.ac.uk}
\affiliation{\addCambridge} 

\author{Jonathan R.\ Gair}
\email{j.gair@ed.ac.uk}
\affiliation{\addCambridge}\affiliation{\addEdinburgh}

\date\today

\begin{abstract}
Folding uncertainty in theoretical models into Bayesian parameter estimation is necessary in order to make reliable inferences. A general means of achieving this is by marginalising over model uncertainty using a prior distribution constructed using Gaussian process regression (GPR). As an example, we apply this technique to the measurement of chirp mass using (simulated) gravitational-wave signals from binary black holes that could be observed using advanced-era gravitational-wave detectors. Unless properly accounted for, uncertainty in the gravitational-wave templates could be the dominant source of error in studies of these systems. We explain our approach in detail and provide proofs of various features of the method, including the limiting behaviour for high signal-to-noise, where systematic model uncertainties dominate over noise errors. We find that the marginalised likelihood constructed via GPR offers a significant improvement in parameter estimation over the standard, uncorrected likelihood both in our simple one-dimensional study, and theoretically in general. We also examine the dependence of the method on the size of training set used in the GPR; on the form of covariance function adopted for the GPR, and on changes to the detector noise power spectral density.
\end{abstract}



\maketitle

\section{Introduction}\label{intro}

The era of advanced ground-based interferometric gravitational-wave (GW) detectors is here. The Advanced LIGO detectors \cite{2010CQGra..27h4006H,2015CQGra..32g4001T} in the USA started observing in September 2015, whilst the Advanced Virgo detector \cite{Acernese2009,2015CQGra..32b4001A} in Europe is expected to come online shortly afterwards \cite{2013arXiv1304.0670L}. The principal target sources of GWs for these detectors are the coalescences of pairs of compact objects, either neutron stars or black holes. For all sources there is great uncertainty in the quoted event rate estimates, at least an order of magnitude in either direction \cite{2010CQGra..27q3001A}, but regardless of the astrophysical uncertainty, the prospect of a first detection is imminent.\footnote{While this paper was in proof, the first detection was announced \cite{2016PhRvL.116f1102A}.}

The detection of GW signals is most efficient when we have accurate waveform models that can be matched to any signals in the (noisy) detector data.\footnote{It is possible to detect GWs without templates by looking for coherent excess power in the detectors, e.g., \cite{2008CQGra..25k4029K,2015CQGra..32m5012C,2015arXiv150906423K}. This is effective for short-duration signals corresponding to high-mass binaries.} For parameter estimation (PE), it is even more important that the template faithfully matches the true signal, as otherwise we could infer biased parameter values. Matching a template to GW data requires that the model waveform remains accurate over the entire duration of the signal; typically of the order of hundreds of seconds for neutron-star binaries and tens of seconds for black-hole binaries in the advanced-detector era (with frequency sensitivity down to $10~\mathrm{Hz}$). Although higher mass sources have shorter waveforms (in the detector band), these present more of a challenge for modelling as they have detectable merger and ringdown components. In contrast, binary neutron stars only have the (easier to model) inspiral part of the waveform in band. In this paper, we are concerned with problems that arise from inaccurate models; therefore, we focus our attention on black-hole binaries where the issue of waveform uncertainty is most acute. However, the techniques we develop could equally be applied to neutron-star binaries or any other uncertain signal. Inaccurate waveform models are known to cause significant systematic errors when recovering source parameters from observations with both ground-based \cite{2001PhRvD..63h2005C} and space-based GW detectors \cite{PhysRevD.76.104018}.

There are two problems that arise when using inaccurate signal models: the \emph{detection} problem, and the \emph{PE} problem. The detection problem is that the inaccurate model does not perfectly match to the physical waveform, leading to a loss of signal-to-noise (SNR) ratio and, hence, a lower chance of detection (for the same false alarm probability). The PE problem, which is the focus of this paper, is that the model waveform which has the best overlap with the physical signal in the data generally has parameter values offset from the true source parameters, leading to a \emph{systematic} error in any parameter estimates.

Recently, some of the authors proposed a novel method of improving the detection and PE prospects of complicated physical phenomena in noisy data \cite{Letter}. The method applies generally to any situation where accurate models of the signal are available, but computational constraints mean that routine detection and PE tasks must be carried out with cheaper, less accurate, models.

In the case of binary black-holes, there are many different physical effects that should be included in waveform models, such as the merger and ringdown phases following the inspiral, the presence of generic spins and precession, eccentricity and higher-order modes. All these phenomena can, in principle, be simulated, thanks to recent rapid progress in numerical relativity (NR) \cite{PhysRevLett.95.121101,2006PhRvL..96k1101C,2006PhRvL..96k1102B}. However, NR simulations are extremely expensive, only a few hundred have been performed to date (see \cite{2013PhRvL.111x1104M,2014CQGra..31k5004A} and references therein), and these typically consist of only the final few tens of orbits (although, see \cite{Szilagyi:2015rwa}). Detection and PE are therefore currently performed using less expensive waveform approximants. The existence of NR waveforms has permitted the calibration of analytic inspiral--merger--ringdown approximants such as the effective-one-body--NR (EOBNR) \cite{1999PhRvD..59h4006B,2000PhRvD..62f4015B,2011PhRvD..84l4052P,2014PhRvD..89f1502T} or IMRPhenom \cite{2010PhRvD..82f4016S} families, with recent efforts concentrating on including the effects of precession in these \cite{2012PhRvD..86j4063S,2014PhRvL.113o1101H,2015PhRvD..91b4043S}.\footnote{See \cite{2013PhRvD..87j4003L} for a study of systematic error (or lack thereof) from using EOBNR waveforms with NR injections.} For some recent PE work with inspiral merger and ringdown waveform models see \cite{2015arXiv150505607G,2015arXiv150305953V,2015PhRvD..92b2002G}. Historically, PE has used models based on the post-Newtonian formalism~\cite{2014LRR....17....2B}, such as the TaylorF2 and TaylorT2 waveforms \cite{PhysRevD.80.084043}. Despite these lacking some of the relevant features, they are sufficiently quick to calculate that they can be simply used in PE algorithms.

To include uncertainty in waveform templates whilst minimising computational expense, we use Gaussian process regression (GPR) to estimate the effects of waveform errors. The method involves constructing a training set of the waveform differences between an expensive, accurate waveform and a cheaper, less accurate waveform. For the accurate waveforms it might be necessary to use some combination of NR and the NR calibrated approximants discussed above, depending on the numbers and length of the available NR simulations. The waveform difference is evaluated at a relatively small number of points in parameter space and stored for later use. GPR is then used to interpolate the difference across parameter space to give a best estimate and a corresponding uncertainty at a general point in parameter space. This interpolation provides a prior probability distribution on the waveform difference which is then used in marginalising the likelihood over waveform uncertainty. The result is an expression for the likelihood in terms of the cheaper waveform model, but with corrections coming from the training set. This marginalised likelihood is negligibly more complicated or computationally expensive to evaluate than the standard expression, but provides a better estimate of the true likelihood surface (and hence the posterior), factoring in our imperfect knowledge of the waveform. Therefore, we have not only built a relatively inexpensive waveform approximant that can include additional physics, but we have also accounted for (marginalised over) the uncertainty in our new approximant. 

If the standard likelihood with an approximate waveform is used for PE then, in general, biased parameter estimates are obtained.\footnote{This is commonly assessed in the GW literature using probability--probability (P--P) plots \cite{2014PhRvD..89h4060S,2015ApJ...804..114B}; for a catalogue of events, these plot the cumulative fraction of events where the true parameter is found within the credible interval corresponding to a given probability. If the posteriors are well calibrated, then a proportion $P$ should fall in the $P$ credible interval, and the plot is a diagonal line. Introducing bias means that the line sags below the optimal diagonal.} It has recently been shown by some of the authors \cite{PPplots} that, under certain conditions, this bias is completely removed by the marginalised likelihood, and, more generally, that the bias is always reduced by the marginalised likelihood.

The technique of GPR assumes that the data in the training set have been drawn from a Gaussian process (GP) on the parameter space with a mean and a covariance function either specified \emph{a priori} or estimated from the training set itself. The interpolation is then achieved by calculating the conditional probability for the GP at some new parameter point given the known training set values, the mean and the covariance. GPR provides a convenient non-parametric way to interpolate the waveform differences, and has the additional advantage that, by construction, it provides a Gaussian probability distribution for the unknown waveform difference which can be analytically marginalised over. This is important because it means no extra nuisance parameters are added to the PE task which would slow down an already expensive process.

The outline of this paper is as follows. In Sec.~\ref{sec:method} the concept of the marginalised likelihood is introduced and the use of GPR in its construction is described in detail; we limit ourselves to interpolation across parameter space and not across frequency. The main choice made in implementing GPR is the specification of the covariance function; Sec.~\ref{sec:cov} discusses how the properties of the covariance function affect the properties of the corresponding GP, and the effects of different choices of covariance function are examined in a toy one-dimensional GPR problem. The marginalised likelihood possesses several properties which make it appealing for GW astronomy; Sec.~\ref{sec:props} presents proofs of these and discussions of their significance. In Sec.~\ref{sec:allres} the implementation of the marginalised likelihood is described for an illustrative one-dimensional example; here, properties of the interpolated waveforms are examined and PE results for the marginalised likelihood are also presented. Additional material on the effect of changing the detector noise properties on the interpolated waveforms are considered in App.~\ref{sec:diffPSD}. Finally, concluding remarks and a discussion of future directions for implementing the marginalised likelihood are presented in Sec.~\ref{sec:discussion}.

\section{The method}\label{sec:method}

In this section we detail how we incorporate waveform uncertainties into GW data analysis. The material presented is an expansion of that in \cite{Letter}. In Sec.~\ref{sec:marginalise} we introduce the standard likelihood function and show how model uncertainties can be treated like nuisance parameters that can be integrated out (marginalised over). Performing this integration requires that a prior probability distribution is specified for the model uncertainties, this is constructed using GPR. This is introduced in Sec.~\ref{sec:GPR}, where we briefly summarise some key results pertaining to GPR; further details can be found in standard textbooks (e.g., \cite{MacKay,GPR,Adler}). The result of the integration is the \emph{marginalised likelihood} presented in Eq.~\eqref{eq:finalresult} which accurately encodes our state of knowledge of the signal parameters, given our imperfect waveform models and the noisy data.

\subsection{The marginalised likelihood}\label{sec:marginalise}

We consider the scenario where we can construct two different waveform models, one accurate but computationally expensive, the other less accurate but quick to calculate. We use the parameters vector $\vec{\lambda}$ to fully characterise the GW signal; Latin indices from the beginning of the alphabet ($a,\,b,\,\ldots$) will be used to label the different components of this vector, and repeated indices should be summed over. The accurate waveforms will be referred to as the exact waveform $h(\vec{\lambda})$, although the method does not require that the accurate waveforms are perfect (see Sec.~\ref{subsec:covnoise}). The cheaper approximate waveform $H(\vec{\lambda})$ is related to $h(\vec{\lambda})$ by the waveform difference
\begin{equation}
\label{eq:waveformerror}
H(\vec{\lambda})=h(\vec{\lambda})+\delta h(\vec{\lambda})\,.
\end{equation}
The waveform templates may be calculated in either the time domain $h(t;\vec{\lambda})$ or the frequency domain $\tilde{h}(f;\vec{\lambda})$; the dependence of the waveform on time or frequency is suppressed in our notation for brevity.

In the context of modelling binary black-hole coalescences there are several highly accurate waveform approximants available, for example, NR waveforms \cite{2014CQGra..31k5004A} or spin EOBNR (SEOBNR) models \cite{2011PhRvD..84l4052P,2012PhRvD..86b4011T,2014PhRvD..89f1502T}. There are also multiple possibilities for the approximate waveform family, for example, the Taylor family of approximants \cite{PhysRevD.80.084043}. For the proof-of-principal numerical calculations in this paper, we need to be able to perform mock PE runs with both waveform families so that we can assess our marginalisation technique does indeed offer a significant improvement. Therefore, we will pick both approximants to be quick to compute, rather than selecting on accuracy: our choice of waveform family is discussed in more detail in Sec.~\ref{sec:model}.

In a PE study, we wish to construct the posterior probability distribution for the signal parameters given the observed data (and any prior information we have about the source) $p(\vec{\lambda}|s)$. From Bayes' theorem, the posterior is given by
\begin{equation}
p(\vec{\lambda}|s) = \frac{L'(s|\vec{\lambda})\pi (\vec{\lambda})}{{\mathcal{Z}}'(s)}\, ,
\end{equation}
where (keeping the notation of \cite{Letter}) $L'(s|\vec{\lambda})$ is the likelihood, $\pi(\vec{\lambda})$ is the prior distribution on the parameters and ${\mathcal{Z}}'(s)$ is the normalising evidence
\begin{equation}\label{eq:evidence}
{\mathcal{Z}}'(s) = \int L'(s|\vec{\lambda}) \pi (\vec{\lambda}) \dd\vec{\lambda}\,.
\end{equation}
In a Bayesian analysis the evidence ${\mathcal{Z}}'(s)$ can be used as the detection statistic (by comparing it with the evidence for the null hypothesis to form the Bayes' factor) \cite{2010PhRvD..81f2003V}, and the positions and widths of peaks in the posterior $p(\vec{\lambda}|s)$ are used to give the parameter estimates and associated uncertainties \cite{2015PhRvD..91d2003V}. For simplicity (although it is not necessary to do so), we assume throughout that $\pi(\vec{\lambda})$ is flat within the relevant region of parameter space. The single remaining challenge is to calculate the likelihood $L'(s|\vec{\lambda})$.

For a detector with stationary, Gaussian noise with power spectral density $S_n(f)$ \cite{2015CQGra..32a5014M}, the likelihood is given by \cite{1994PhRvD..49.2658C}
\begin{equation}\label{eq:Lexact}
L'(s|\vec{\lambda})\propto \exp\left( -\frac{1}{2} \left\langle  s- h(\vec{\lambda})\middle| s- h(\vec{\lambda}) \right\rangle \right) \, .
\end{equation}
Here the noise-weighted inner product has been defined as \cite{1992PhRvD..46.5236F}
\begin{eqnarray}
\left\langle x\middle|y\right\rangle &=& 4\Re\left\{ \int_{0}^{\infty}\dd f\,\frac{\tilde{x}(f)\tilde{y}(f)^{*}}{S_{n}(f)} \right\} \nonumber \\
&=&4\Re\left\{ \sum_{\kappa=1}^{M}\delta f\,\frac{\tilde{x}(f_{\kappa})\tilde{y}(f_{\kappa})^{*}}{S_{n}(f_{\kappa})} \right\}\, ,
\label{eq:innerprod}
\end{eqnarray}
where $\kappa$ labels the $M$ frequency bins with resolution $\delta f$. We define the norm of a waveform as
\begin{equation}
\left\|x\right\| = \sqrt{\left\langle x\middle|x\right\rangle}\,,
\label{eq:norm}
\end{equation}
for a signal this is equivalent to its SNR.

In practice it can be unfeasible to sample from the likelihood distribution in Eq.~\eqref{eq:Lexact} because it is prohibitively expensive to calculate the exact waveforms $h(\vec{\lambda})$; instead, we must reply on the approximate waveforms to calculate an approximate likelihood,
\begin{equation}\label{eq:Lapprox}
L(s|\vec{\lambda})\propto \exp\left( -\frac{1}{2} \left\| s- H(\vec{\lambda}) \right\|^2 \right)\,.
\end{equation}
For a good approximant
\begin{equation}
 L(s|\vec{\lambda}) \approx L'(s|\vec{\lambda}) \, ;
\end{equation}
the natural way to improve this agreement is to construct (inevitably more expensive) approximants that have smaller waveform differences $\delta h(\vec{\lambda})$. Instead, the proposal of this paper is to replace $L(s|\vec{\lambda})$ with a new likelihood which accounts for the uncertainty in the waveforms. The alternative likelihood is
\begin{widetext}
\begin{eqnarray}
\label{eq:Lcal}
{\mathcal{L}}(s|\vec{\lambda}) \propto  \int \dd\left[\delta h(\vec{\lambda})\right]\, P[\delta h(\vec{\lambda})] \exp\left( -\frac{1}{2} \left\|  s- H(\vec{\lambda})+\delta h(\vec{\lambda}) \right\|^2 \right) \, .
\end{eqnarray}
\end{widetext}
This new likelihood has marginalised over the uncertainty in the waveform difference using the (as yet unspecified) prior on the waveform difference $P[\delta h(\vec{\lambda})]$.

The prior on the waveform difference should include the information available from the limited number of available accurate waveforms and could also encode our prior expectations about the signal, for example, that the approximate waveforms are most accurate at early times (or equivalently at low frequencies) when the orbiting bodies are well separated \cite{2014LRR....17....2B}, but gradually become inaccurate as the bodies inspiral. At most points in parameter space, an accurate waveform is not available, and so it is necessary to interpolate the waveform difference across parameter space while simultaneously accounting for the error this introduces. It would seem that the problem rapidly becomes complicated, and even if a suitable prior could be constructed the computational time needed to evaluate ${\mathcal{L}}(s|\vec{\lambda})$ would make it impractical in most contexts.

Fortunately, the technique of GPR provides a natural way to interpolate the waveform differences across parameter space, incorporating all necessary prior information. GPR also has the additional property that it naturally returns an expression for $P[\delta h(\vec{\lambda})]$ which is a Gaussian in $\delta h (\vec{\lambda})$. Since the exponential factor in Eq.~\eqref{eq:Lcal} is also Gaussian in $\delta h (\vec{\lambda})$, the functional integral can be evaluated analytically. This gives an analytic expression for ${\mathcal{L}}(s|\vec{\lambda})$ which can be evaluated in approximately the same computational time as $L(s|\vec{\lambda})$.

Henceforth, for brevity, the $s$ dependence will be suppressed in all likelihoods, i.e.\ $L'(\vec{\lambda})\equiv L'(s|\vec{\lambda})$, $L(\vec{\lambda})\equiv L(s|\vec{\lambda})$, and ${\mathcal{L}}(\vec{\lambda})\equiv {\mathcal{L}}(s|\vec{\lambda})$.

\subsection{Gaussian process regression}\label{sec:GPR}

Assume that we have access to accurate waveforms at a few values of the parameters $\{h(\vec{\lambda}_{i})\mid i=1,\,2,\ldots ,\,N\}$ and can cheaply compute approximate waveforms at the same parameter values. Our training set is the set of waveform differences
\begin{equation}
{\mathcal{D}} = \left\{\left(\vec{\lambda}_{i},\,\delta h (\vec{\lambda}_{i})\right) \,\middle|\, i =1,\,2,\ldots,\,N \right\} \, ,
\end{equation}
where necessary the Latin indices $i,\,j,\,\ldots$ will be used to label the different components of the training set (repeated indices are not summed over unless specified). It is now necessary to interpolate the training set to obtain the prior on the waveform difference first defined in Eq.~\eqref{eq:Lcal},
\begin{equation}
P[\delta h] \equiv P(\delta h (\vec{\lambda})|{\mathcal{D}},\,{\mathcal{I}}) \,,
\end{equation}
where ${\mathcal{I}}$ is any other prior information we possess about the waveforms. The simplest and most natural choice for such a prior is to assume that the waveform difference is a realisation of a GP (a Gaussian is the the maximum-entropy distribution given that we know a characteristic range of variation \cite{Shannon1948a}),
\begin{equation}\label{eq:deltahGP}
\delta h(\vec{\lambda}) \sim {\mathcal{GP}}(m(\vec{\lambda}),\,k(\vec{\lambda},\vec{\lambda}')) 
\end{equation} 
A GP can loosely be thought of as the generalisation of a Gaussian distribution to an infinite number of degrees of freedom. It is completely specified by the mean $m(\vec{\lambda})$ and covariance $k(\vec{\lambda},\vec{\lambda}')$ functions in the same way as a Gaussian distribution is fully specified by a mean and variance. More formally, a GP is an infinite collection of variables, any finite subset of which are distributed as a multivariate Gaussian. For a set of parameter points $\{\vec{\lambda}_{i}\}$, including, but not limited to, the training set ${\mathcal{D}}$, 
\begin{equation}\label{eq:tempprob}
 \left[\delta h (\vec{\lambda}_{i})\right]\sim {\mathcal{N}}({\boldsymbol{m}},\,{\boldsymbol{K}})\,,
\end{equation}
where the mean vector and covariance matrix of this Gaussian distribution are fixed by the corresponding functions of the GP,
\begin{equation}\label{eq:covmatrix}
\left[{\boldsymbol{m}}\right]_{i} = m(\vec{\lambda}_{i})\,,\quad \left[{\boldsymbol{K}}\right]_{ij} = k(\vec{\lambda}_{i},\vec{\lambda}_{j})\, ,
\end{equation}
with probability density function (here correcting the normalising prefactor written in~\cite{Letter} which mistakenly included a square root)
\begin{widetext}
\begin{equation} \label{eq:tempprob2}
P\left(\left\{\delta h (\vec{\lambda}_{i})\right\}\right) = \frac{1}{(2\pi)^{N }\left|{\boldsymbol{K}}\right|} \exp\left(-\frac{1}{2}\sum_{i,\,j}\left[{\boldsymbol{K}}^{-1}\right]_{ij}\left(\delta h (\vec{\lambda}_{i})\middle|\delta h (\vec{\lambda}_{j})\right)\right)\,.
\end{equation}
\end{widetext}
The round brackets denote a new inner product with respect to some noise weighting $S'_{n}(f)$, which we leave unspecified for the moment;
\begin{eqnarray}
\left(x|y\right)&=&4\Re\left\{\int_{0}^{\infty}\mathrm{d}f\,\frac{\tilde{x}(f)\tilde{y}(f)^{*}}{S'_{n}(f)}\right\} \nonumber \\
&=&4\Re\left\{\sum_{\kappa=1}^{M}\delta f\,\frac{\tilde{x}(f_{\kappa})\tilde{y}(f_{\kappa})^{*}}{S'_{n}(f_{\kappa})}\right\}\,.
\end{eqnarray}

In writing down Eq.~\eqref{eq:tempprob2} and stipulating that the covariance function $k(\vec{\lambda},\vec{\lambda}')$ has no dependence on frequency, we are effectively assuming that a) the parameter space structure of the model errors is frequency independent; and b) the typical size of errors has a frequency dependence proportional to $\sqrt{S'_{n}(f)}$. Under the assumption that waveform model errors are uncorrelated in frequency, the normalising factor in Eq.~\ref{eq:tempprob2} should be raised to the power $M$; however, this assumption leads to model errors that average to zero over frequency and have only a small effect on the likelihood. The optimal means of incorporating frequency dependence would be to introduce an additional covariance function in frequency as well as the covariance in parameter space. This frequency covariance introduces a correlation length scale in frequency which can be learnt from the training set in exactly the same manner as we describe below for correlations in $\vec{\lambda}$. This correlation length scale reduces the number of independent frequencies from $M$ to some new effective number $M_\mathrm{eff}$.

Performing this double GPR interpolation in $f$ and $\vec{\lambda}$ is beyond the scope of the current paper. Instead, here we are in effect setting $M_\mathrm{eff}=1$, giving the expression in Eq.~\ref{eq:tempprob2}; this is analogous to assuming that all the frequency bins of the noise-weighted waveform at a particular point in parameter space are perfectly correlated. Setting $M_\mathrm{eff}=1$ gives the largest uncertainty of any fixed number of independent frequencies and is therefore a conservative choice. Despite these simplifications, our marginalised likelihood has many desirable properties (which we discuss and prove in Sec.~\ref{sec:props}), and performs well in the numerical example presented in Sec.~\ref{sec:allres}. We will return to the more general problem of performing the extended GPR including frequency in the future.

Specifying how we compute the mean and variance for the GP determines how the waveforms are interpolated and fixes our prior for waveform uncertainty across parameter space. Our GP has a zero mean as we have chosen to interpolate the waveform difference rather than the waveform directly. By first subtracting off an approximate model we leave a quantity which is uncertain, but has no known bias. If we had some additional prior knowledge that the approximate waveform was systematically wrong across parameter space, then this should be added into the approximate model so that the zero-mean assumption becomes valid. Identical results for the marginalised likelihood could also be obtained by directly interpolating the accurate waveforms using a GP with a mean equal to the approximate waveforms; however, we choose to interpolate waveform differences because zero-mean GPs are simpler to handle numerically.

Specifying the covariance function is central to GPR as it encodes our prior expectations about the properties of the function being interpolated. Possibly the simplest and most widely used choice for the covariance function is the squared exponential (SE) \cite{GPR}
\begin{equation}\label{eq:covfunc}
k(\vec{\lambda}_{i},\vec{\lambda}_{j}) = \sigma_{f}^{2}\exp\left[-\frac{1}{2}g_{ab}(\vec{\lambda}_{i}-\vec{\lambda}_{j})^{a}(\vec{\lambda}_{i}-\vec{\lambda}_{j})^{b}\right] \,,
\end{equation}
which defines a stationary, smooth GP. In Eq.~\eqref{eq:covfunc}, a scale $\sigma_{f}$ and a (constant) metric $g_{ab}$ for defining a modulus in parameter space have been defined. These are called \emph{hyperparameters} and we denote them as $\vec{\theta} =\left\{\sigma_{f},g_{ab}\right\}$, with Greek indices $\mu,\,\nu,\,\ldots$ to label the components of this vector. If the available accurate waveforms contain some uncertainty then this can also be included by adding a diagonal matrix ${\boldsymbol{C}}$ to Eq.~\eqref{eq:covfunc}, where the element $C_{ii}$ (no summation) is the uncertainty in the accurate simulation at $\vec{\lambda}_i$; this is discussed further in Sec.~\ref{subsec:covnoise}.

The probability in Eq.~\eqref{eq:tempprob2} is referred to as the \emph{hyperlikelihood}, or alternatively the \emph{evidence} (as in \cite{Letter}) for the training set; it is the probability that that particular realisation of waveform differences was obtained from a GP with a zero mean and specified covariance function. The hyperlikelihood depends only on the hyperparameters and the quantities in the training set, so we denote it as $Z(\vec{\theta}|{\mathcal{D}})$. The log hyperlikelihood is
\begin{eqnarray}
\ln Z(\vec{\theta}|{\mathcal{D}}) & = &-\frac{N}{2}\ln(2\pi) \nonumber\\*
 && -\frac{1}{2}\sum_{i,\,j}\inv\left[k(\vec{\lambda}_{i},\vec{\lambda}_{j})\right]\left(\delta h (\vec{\lambda}_{i})\middle| \delta h (\vec{\lambda}_{j})\right)\nonumber \\*
 && -\frac{1}{2}\ln \left| \det\left[k(\vec{\lambda}_{i},\vec{\lambda}_{j})\right] \right| \, . \label{eq:evtrainset}
\end{eqnarray}

For all subsequent calculations the values of the hyperparameters are fixed to their optimum values $\vec{\theta}_{\mathrm{op}}$, defined as those which maximise the hyperlikelihood:
\begin{equation}
\left.\frac{\partial Z(\vec{\theta}|{\mathcal{D}})}{\partial \theta^{\mu}}\right|_{\vec{\theta}\,=\,\vec{\theta}_{\mathrm{op}}} = 0 \,.
\end{equation}
Maximising the hyperlikelihood with respect to $\vec{\theta}$ is one of many approaches which could be taken. For example, a better motivated approach would be to consider the hyperparameters as nuisance parameters in addition to the source parameters $\vec{\lambda}$, and marginalise over them while sampling an expanded likelihood,
\begin{eqnarray}
\Lambda_{\mathrm{expanded}}(\vec{\lambda},\vec{\theta}|{\mathcal{D}}) &\propto {\mathcal{L}}(\vec{\lambda}|\vec{\theta},{\mathcal{D}})Z(\vec{\theta}|{\mathcal{D}}).
\end{eqnarray}
The disadvantage of this approach is that the hyperlikelihood is \emph{much} more expense to compute than the standard approximate likelihood and the inclusion of extra nuisance parameters also slows down any PE. In contrast, our proposed method of maximising the likelihood is a convenient heuristic which is widely used in other contexts \citep{MacKay1999,snelson2005sparse,quinonero2007approximation} and allows all the additional computation to be done offline. It would be useful, in future work, to check explicitly that the different ways of dealing with the hyperparameters give consistent results in the context of GW source modelling.

Having fixed the properties of the covariance function by examining the training set, we can now move on to using the GP as a predictive tool. The defining property of the GP is that \emph{any} finite collection of variables drawn from it is distributed as a multivariate Gaussian in the manner of Eq.~\eqref{eq:tempprob2}. Therefore, the set of variables formed by the training set plus the waveform difference at one extra parameter point $\delta h (\vec{\lambda})$ is distributed as
\begin{equation}\label{eq:GPRdistDplusextra}
\left[\begin{matrix}\delta h(\vec{\lambda}_{i})\\\delta h(\vec{\lambda})\end{matrix}\right] \sim {\mathcal{N}}\left({\boldsymbol{0}},{\boldsymbol{\Sigma}}\right)\,,\quad {\boldsymbol{\Sigma}} =\left(\begin{matrix}{\boldsymbol{K}}&{\boldsymbol{K}}_{*}\\{\boldsymbol{K}}_{*}^{\mathrm{T}}&K_{**}\end{matrix}\right)\,,
\end{equation}
where ${\boldsymbol{K}}$ is defined in Eq.~\eqref{eq:covmatrix} and the vector ${\boldsymbol{K}}_{*}$ and scalar $K_{**}$ are defined as
\begin{equation}
\left[{\boldsymbol{K}}_{*}\right]_{i} = k(\vec{\lambda}_{i},\vec{\lambda})\,, \quad K_{**}=k(\vec{\lambda},\vec{\lambda}) \,.
\label{eq:vecandscalar}
\end{equation}
On the right-hand side of Eq.~\eqref{eq:GPRdistDplusextra} all the quantities are known because the hyperparameters have been fixed to their optimum values, and on the left hand side all the quantities are known (from the training set) except for $\delta h (\vec{\lambda})$. Therefore, the conditional probability of the unknown waveform difference given the known differences in ${\mathcal{D}}$ can be found. This conditional probability is given by (e.g., \cite{MacKay,GPR})
\begin{widetext}
\begin{equation}\label{eq:GRPprior}
P[\delta h(\vec{\lambda})]=\frac{1}{{2\pi\sigma^{2}(\vec{\lambda})\prod_{\kappa=1}^{M} S'_{n}(f_{\kappa})}} \exp\left(-\frac{\left(\delta h(\vec{\lambda})-\mu(\lambda) \middle| \delta h(\vec{\lambda})-\mu(\lambda) \right)}{2\sigma^{2}(\vec{\lambda})}\right) \, ,
\end{equation}
\end{widetext}
where the GPR mean and its associated error have been defined as
\begin{eqnarray}
\mu(\vec{\lambda})&=& \sum_{i,\,j}\left[{\boldsymbol{K}}_{*}\right]_{i} \left[{\boldsymbol{K}}^{-1}\right]_{ij}\delta h(\vec{\lambda}_{j}) \, , \label{eq:GPRmeanequ}\\
\sigma^{2}(\vec{\lambda})&=& K_{**}-\sum_{i,\,j}\left[{\boldsymbol{K}}_{*}\right]_{i} \left[{\boldsymbol{K}}^{-1}\right]_{ij}\left[{\boldsymbol{K}}_{*}\right]_{j}\, . \label{eq:GPRsigma2}
\end{eqnarray}

Furnished with the expression for $P[\delta h(\vec{\lambda})]$, the marginalised likelihood in Eq.~\eqref{eq:Lcal} can now be evaluated. The integrand in Eq.~\eqref{eq:Lcal} is the product of two Gaussians and can be calculated analytically,
\begin{widetext}
\begin{eqnarray}\label{eq:finalresultprime}
{\mathcal{L}}(\vec{\lambda}) &\propto& \frac{1}{{1+\sigma^{2}(\vec{\lambda})\prod_{\kappa=1}^{M}\left(S'_{n}(f_{\kappa})/S_{n}(f_{\kappa})\right)}} \exp\left(-\frac{1}{2}\left[s-H(\vec{\lambda})+\mu(\vec{\lambda}) \middle| s-H(\vec{\lambda})+\mu(\vec{\lambda})\right]\right).
\end{eqnarray}
\end{widetext}
The square brackets denote a third inner product with respect to the new noise weighting $S''_{n}(f)$,  where $S''_{n}(f,\vec{\lambda}) \equiv S_{n}(f)+\sigma^{2}(\vec{\lambda})S'_{n}(f)$,
\begin{eqnarray}
\left[x|y\right]&=&4\Re\left\{\int_{0}^{\infty}\mathrm{d}f\,\frac{\tilde{x}(f)\tilde{y}(f)^{*}}{S''_{n}(f)}\right\} \nonumber \\
 &=& 4\Re\left\{\sum_{\kappa=1}^{M}\delta f\,\frac{\tilde{x}(f_{\kappa})\tilde{y}(f_{\kappa})^{*}}{S''_{n}(f_{\kappa})}\right\} \,.
\end{eqnarray}
For the remainder of this paper, for simplicity, we take $S'_{n}(f)=S_{n}(f)$ so the three signal inner products we have defined become ${\langle \cdot|\cdot\rangle=(\cdot|\cdot)=[\cdot|\cdot]/(1+\sigma^{2}(\vec{\lambda}))}$ \cite{Letter}. With this simplifying assumption, the marginalised likelihood becomes
\begin{eqnarray}\label{eq:finalresult}
{\mathcal{L}}(\vec{\lambda}) &\propto& \frac{1}{{1+\sigma^{2}(\vec{\lambda})}} \nonumber \\
 & &  \times \exp\left(-\frac{1}{2}\frac{\left\| s-H(\vec{\lambda})+\mu(\vec{\lambda})\right\|^2}{1+\sigma^{2}(\vec{\lambda})}\right).
\end{eqnarray}
As mentioned earlier Eq.~\eqref{eq:tempprob2} issues that the waveform model errors are uncorrelated in frequency. The assumption that $S'_{n}(f)=S_{n}(f)$ additionally assumes that the typical size of the waveform error at a frequency $f$ is given by $\sqrt{S_n(f)}$. This choice can be motivated to a certain extent by examining the hyperlikelihood in Eq.~\eqref{eq:evtrainset} which is used to train the GP. This hyperlikelihood contains the overlap matrix ${(\delta h(\vec{\lambda}_{i})|\delta h(\vec{\lambda}_{j}))}$. Choosing $S'_{n}(f)=S_{n}(f)$ acts to downweight the correlations at frequencies we are insensitive to (ignoring errors we cannot measure) and hence the resulting hyperparameters give an interpolant which is tuned to better represent the waveform correlations at the frequencies to which we are most sensitive: we weight waveform errors based upon their impact on the likelihood. The assumption of frequency-independent models errors gives a value for the GPR uncertainty $\sigma^{2}$ in Eq.~\eqref{eq:finalresult} that is also frequency-independent. This can be shown to be a conservative choice in the sense that it gives broader and less informative posteriors.

In a follow-on study we will provide a proof of the conservative nature of this assumption and examine a number of different choices for the weighting function $S'_{n}(f)$, but we use the simplifying assumption $S'_{n}(f)=S_{n}(f)$ throughout the current paper. Despite these simplifying assumptions, we find that the resulting likelihood in Eq.~\eqref{eq:finalresult} performs well.  In App.~\ref{sec:diffPSD} we examine the sensitivity of the method to small changes in the noise curve $S_{n}(f)$ which will occur in real experiments.

In Eq.~\eqref{eq:finalresult} the best fit waveform has shifted by an amount $\mu(\vec{\lambda})$; this is the best estimate of the waveform difference returned by the GPR. The quantity $H(\vec{\lambda})+\mu(\vec{\lambda})$ can be regarded as a new waveform approximant built from the accurate and approximate waveforms with the aid of GPR. However, a bonus of this way of including the training set directly into the likelihood is that the extra uncertainty associated with using the GPR as an interpolant is automatically included via the broadening of the posterior caused by $\sigma^{2}(\vec{\lambda})\ge 0$. 

In this section we have explained how uncertainty in waveform models can be included in PE through use of a marginalised likelihood. We defined such a likelihood in Eq.~\eqref{eq:Lcal}, but the marginalisation requires a prior probability on the waveform uncertainty across parameter space. We construct this from a training set using GPR; the resulting prior is given in Eq.~\eqref{eq:GRPprior}. Since this is of Gaussian form, we can marginalise analytically to produce the new likelihood Eq.~\eqref{eq:finalresult}. The properties of this marginalised likelihood are explored extensively throughout the remainder of this paper.

In Secs.~\ref{sec:cov} and \ref{sec:props} we discuss theoretical properties of the GPR and marginalised likelihood respectively. A reader who is primarily interested in the PE results obtained with the likelihood in Eq.~\eqref{eq:finalresult} may skip to Sec.~\ref{sec:allres}.

\section{The covariance function}\label{sec:cov}

In the previous section we described how waveform uncertainties could be marginalised out using a prior constructed by using GPR on a training set. The only aspect of this that is not prescribed by the training data is the choice of the covariance function. This plays an important role in determining the properties of a GP. In this section, we discuss the properties of different choices of the covariance function in GPR. The properties of the covariance functions discussed in this section are known in the GPR literature, but are included here as they are not a widely appreciated in the GW community. The material presented in this section on the covariance function will be used in the interpretation of our results in Sec.~\ref{sec:props} and Sec.~\ref{sec:allres}.

The only necessary requirements we have of a covariance function are that it is a positive definite; i.e.\ for \emph{any} choice of points $\small\{\vec{\lambda}_{i}\small\}$ the covariance matrix $K_{ij}=k(\vec{\lambda}_{i},\vec{\lambda}_{j})$ is positive definite. 

Throughout this paper, GPs are assumed to have zero mean, and therefore be fully specified by the covariance function $k(\vec{\lambda}_{1},\vec{\lambda}_{2})$. However, the proofs regarding continuity and differentiability of GPs discussed in this section, and proved in App.~\ref{app:A}, are done without recourse to the zero-mean assumption. The covariance encodes all information available about the properties of the function being interpolated by the GPR. It is central to the GPR and hence also to the marginalised likelihood.

The covariance function (and the corresponding GP) is said to be \emph{stationary} if the covariance is a function only of $\vec{\tau}= \vec{\lambda}_{1}-\vec{\lambda}_{2}$, furthermore it is said to be \emph{isotropic} if it is a function only of $\tau \equiv |\vec{\tau}| = |\vec{\lambda}_{1}-\vec{\lambda}_{2}|$.\footnote{We have yet to define a metric on parameter space with which to take the norm of this vector (see Sec.~\ref{subsec:metric}), but all that is required here is that a suitably smooth metric exists.} Isotropy of a GP implies stationarity. All of the GPs used for numerical calculations in this paper are isotropic (and hence stationary) $k(\vec{\lambda}_{1},\vec{\lambda}_{2}) \equiv k(\vec{\tau}) \equiv k(\tau)$, although the generalisation to non-stationary GPs is briefly discussed in Sec.~\ref{subsec:distance}.

An example of how the properties of the covariance function relate to the properties of the GP, and hence the properties of the resulting interpolant, is given by considering the \emph{mean-square} (MS) continuity and differentiability of GPs. It can be shown that the first $n_{d}$ MS derivatives of a GP are MS continuous (the GP is said to be $n_{d}$-times MS differentiable) if and only if the first $2n_{d}$ derivatives of the covariance function are continuous at the diagonal point ${\vec{\lambda}_{1}=\vec{\lambda}_{2}=\vec{\lambda}_{*}}$. For a stationary GP this condition reduces to checking the $2n_{d}$ derivatives of $k(\vec{\tau})$ at $\vec{\tau}=\vec{0}$, and for an isotropic GP checking the $2n_{d}$ derivatives of $k(\tau)$ at $\tau=0$. A proof of this result, following \cite{Adler}, is given in App.~\ref{app:A}. It is the smoothness properties of the covariance function at the origin that determine the differentiability of the GP. This result is used in Sec.~\ref{subsec:distance} when discussing different functional forms of covariance for use in GPR.

In this section, the effect of the choice of covariance function on the GPR are explored. We consider three aspects that enter the definition of the covariance function:
\begin{enumerate}[label=(\Alph*)]
\item specifying the distance metric in parameter space $g_{ab}$;
\item specifying the functional form of the covariance with distance $k(\tau)$,
\item and whether or not to include errors $\sigma_{n}$ on the training set points.
\end{enumerate}

Stages A and B cannot be completely separated; there exists an arbitrary scaling, $\alpha$ of the distance $\tau\rightarrow\alpha\tau$ which can be absorbed into the definition of the covariance, $k(\tau)\rightarrow k(\tau/\alpha)$. However, provided the steps are tackled in order, there is no ambiguity.

\subsection{The metric $g_{ab}$}\label{subsec:metric}

The first stage involves defining a distance $\tau$ between two points in parameter space. One simple way of doing this, and the way used in the SE covariance function in Eq.~\eqref{eq:covfunc}, is to define ${\tau^{2} = g_{ab}(\vec{\lambda}_{1}-\vec{\lambda}_{2})^{a}(\vec{\lambda}_{1}-\vec{\lambda}_{2})^{b}}$, where $g_{ab}$ are constant hyperparameters. This distance is obviously invariant under a simultaneous translation of $\vec{\lambda}_{1}\rightarrow\vec{\lambda}_{1}+\vec{\Delta}$ and $\vec{\lambda}_{2}\rightarrow\vec{\lambda}_{2}+\vec{\Delta}$; therefore, this defines a stationary GP. For a $D$-dimensional parameter space, this involves specifying $D(D+1)/2$ hyperparameters $g_{ab}$.

More complicated distance metrics (with a larger number of hyperparameters) are possible if the condition of stationarity is relaxed, i.e.\ $g_{ab}\rightarrow g_{ab}(\vec{\lambda})$. It was demonstrated by \cite{Paci:Sche:2004} how, given a family of stationary covariance functions, a non-stationary generalisation can be constructed. A stationary covariance function can be considered as a kernel function centred at $\vec{\lambda}_{1}$; $k(\vec{\lambda}_{1},\vec{\lambda}_{2})\equiv k_{\vec{\lambda}_{1}}(\vec{\lambda}_{2})$. Allowing a different kernel function to be defined at each point $\vec{\lambda}_{1}$, a new, non-stationary covariance function is $k(\vec{\lambda}_{1},\vec{\lambda}_{2})=\int\mathrm{d}\vec{u}\,k_{\vec{u}}(\vec{\lambda_{1}})k_{\vec{u}}(\vec{\lambda}_{2})$.\footnote{To see that $k$ is a valid covariance function consider an arbitrary series of points $\small\{\vec{\lambda}_{i}\small\}$, and the sum over training set points $I= \sum_{i,j}a_{i}a_{j}k(\vec{\lambda}_{i},\vec{\lambda}_{j})$; for $k$ to be a valid covariance it is both necessary and sufficient that $I\geq 0$. Using the definition of $k$ gives $I=\int\mathrm{d}\vec{u}\, \sum_{i,j} a_{i}a_{j}k_{\vec{u}}(\vec{\lambda_{i}})k_{\vec{u}}(\vec{\lambda}_{j}) = \int\mathrm{d}\vec{u}\,(\sum_ia_{i}k_{\vec{u}}(\vec{\lambda_{i}}))^{2} \geq 0$.}
Applying this procedure to a $D$-dimensional SE function generates a non-stationary analogue \cite{Paci:Sche:2004}
\begin{eqnarray}
k(\vec{\lambda}_{i},\vec{\lambda}_{j}) & = & \sigma_{f}\left|{\mathcal{G}}^{i}\right|^{1/4}\left|{\mathcal{G}}^{j}\right|^{1/4}\left|\frac{{\mathcal{G}}^{i}+{\mathcal{G}}^{j}}{2}\right|^{-1/2} \nonumber \\
 & & \times \exp\left(-\frac{1}{2}\mathcal{Q}_{ij}\right) \,,
\label{eq:non-stationarySE}
\end{eqnarray}
where
\begin{eqnarray}
\mathcal{Q}_{ij} &=& (\vec{\lambda}_{i}-\vec{\lambda}_{j})^{a}(\vec{\lambda}_{i}-\vec{\lambda}_{j})^{b}\left(\frac{{\mathcal{G}}_{ab}^{i}+{\mathcal{G}}_{ab}^{j}}{2}\right)^{-1} ,
\label{eq:Qdef}
\end{eqnarray}
and ${\mathcal{G}}^{i}_{ab} = \inv[g_{ab}(\vec{\lambda}_{i})]$ is the inverse of the parameter-space metric at position $\vec{\lambda}_{i}$. Provided that the metric $g_{ab}(\vec{\lambda})$ is smoothly parameterised this non-stationary SE function retains the smoothness properties discussed earlier.

For the interpolation of waveform differences, it is easy to imagine the potential benefits of using non-stationary GPs. For example, in the case of the spin parameter, it could be imagined that the waveform difference considered as a function of the effective spin of the compact objects $\delta h(\chi)$ would vary on long length scales in $\chi$ for small values of the spin, but on much shorter scales for larger values of the spin. 

The generalisation in Eq.~\eqref{eq:non-stationarySE} involves the inclusion of a large set of additional hyperparameters to characterise how the metric changes over parameter space; for example one possible parameterisation would be the Taylor series
\begin{equation}
g_{ab}(\vec{\lambda})=g_{ab}(\vec{\lambda}_{0})+ \left(\vec{\lambda}^{c}-\vec{\lambda}^{c}_{0}\right)\left.\frac{\partial g_{ab}(\vec{\lambda})}{\partial \lambda^{c}}\right|_{\vec{\lambda}=\vec{\lambda}_{0}}+\ldots \end{equation}
with the hyperparameters $g_{ab}(\vec{\lambda}_{0})$, $\partial g_{ab}(\vec{\lambda})/\partial \lambda^{c}$, and so on. As we see below, the inclusion of even a single extra hyperparameter can incur a significant Occam penalty \cite{MacKay} which pushes the training set to favour a simpler choice of covariance function. For this reason we only consider stationary GPs. However, the generalisation to a non-stationary GP (perhaps in only a limited number of parameters, e.g., spin) should be investigated further in the future. In making this generalisation, one would have to be guided significantly by the prior expectations of which parameters to include and how to parameterise the varying metric.

An alternative to considering non-stationary metrics is instead to try and find new coordinates $\tilde{\lambda} \equiv \tilde{\lambda}(\vec{\lambda})$ such that the metric in these coordinates becomes (approximately) stationary. There could be hope for this approach, as a similar problem has been tackled in the context of template placement for GW searches \cite{1996PhRvD..53.6749O}. Here the problem is to find coordinates such that waveform templates placed on a regular grid in these coordinates have a constant overlap with each other. The waveform match can be viewed as defining a metric in parameter space, and hence the desired coordinates make this metric stationary. For a post-Newtonian inspiral signal, a set of chirp-time coordinates were proposed by \cite{PhysRevD.44.3819} which make the metric nearly stationary. Metrics have also been calculated for inspiral--merger--ringdown models, for example IMRPhenomB \cite{2015PhRvD..91l4042K}. While it could be possible to adapt the parameter-space metrics already calculated for different approximants for use in template placement algorithms to help in constructing our GPR training sets, we do not consider this approach further here.

Throughout the remainder of this paper the metric components $g_{ab}$ are treated as constant hyperparameters fixed to their optimum values, as discussed in Sec.~\ref{sec:method}.

\begin{figure*}
 \centering
 \includegraphics[trim=2cm 0.5cm 2cm 0.5cm, width=0.95\textwidth]{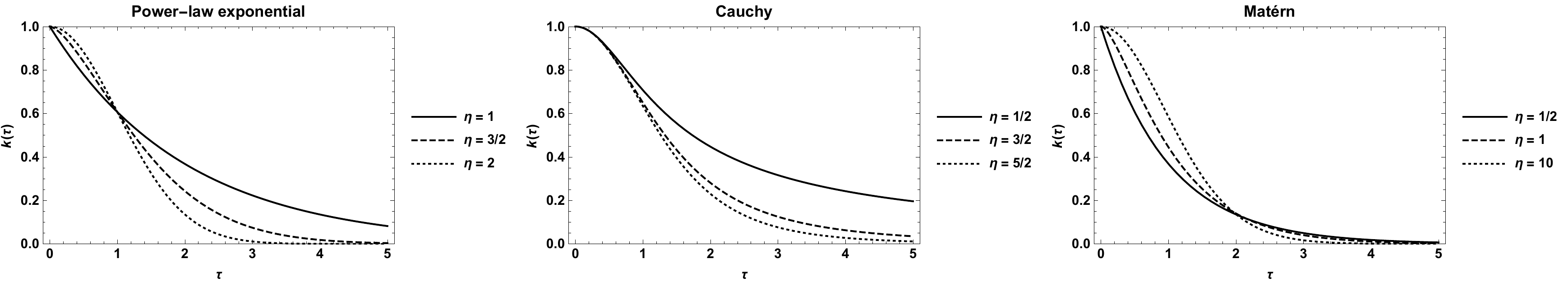}
 \caption{Plots of the different generalisations of the SE covariance function discussed in Sec.~\ref{subsec:distance}. The left-hand panel shows the PLE function, the centre panel shows the Cauchy function, and the right-hand panel shows the Mat\'{e}rn function; in all cases the value of $\sigma_{f}$ was fixed to unity. In each panel the effect of varying the additional hyperparameter is shown by the three curves. For the PLE covariance the case $\eta=2$ recovers the SE covariance, while for the Cauchy and Mat\'ern covariances the case $\eta\rightarrow\infty$ recovers the SE covariance.}
 \label{fig:covfunctions}
\end{figure*}

\subsection{The functional form of $k(\tau)$}\label{subsec:distance}

The second stage of specifying the covariance function involves choosing the function of distance $k(\tau)$. In general whether a particular function $k(\tau)$ is positive definite (and hence is a valid covariance function) depends on the dimensionality $D$ of the underlying space (i.e. $\vec{\lambda}\in \mathbb{R}^{D}$); however, all the functions considered in this section are positive definite for all $D$. Several choices for $k(\tau)$ are particularly common in the literature. These include the SE covariance function (which has already been introduced), given by
\begin{equation}\label{eq:covSE}
k_{\mathrm{SE}}(\tau) =\sigma_{f}^{2}\exp\left(-\frac{1}{2}\tau^{2}\right) \,.
\end{equation}
The \emph{power-law exponential} (PLE) covariance function is given by
\begin{equation}\label{eq:covPL}
k_{\mathrm{PLE}}(\tau)=\sigma_{f}^{2}\exp\left(-\frac{1}{2}\tau^{\eta}\right)  \,,
\end{equation}
where $0<\eta\leq 2$. The PLE reduces to the SE in the case $\eta=2$. The \emph{Cauchy} function is given by
\begin{equation}\label{eq:covCauchy}
k_{\mathrm{Cauchy}}(\tau) =\frac{\sigma_{f}^{2}}{\left(1+\tau^{2}/2\eta\right)^{\eta}}  \,,
\end{equation}
where $\eta>0$. This recovers the SE function in the limit $\eta\rightarrow\infty$. And finally, the \emph{Mat\'{e}rn} covariance function is given by \cite{Stein1999}
\begin{equation}\label{eq:covMatern}
k_{\mathrm{Mat}}(\tau) =\frac{\sigma_{f}^{2}2^{1 - \eta}}{\Gamma (\eta)}\left(\sqrt{2\eta}\,\tau\right)^{\eta}K_{\eta}\left(\sqrt{2\eta}\,\tau\right)  \,,
\end{equation}
where $\eta>1/2$, and $K_{\eta}$ is the modified Bessel function of the second kind \cite{Watson1995}. In the limit $\eta\rightarrow\infty$, the Mat\'{e}rn covariance function also tends to the SE.

Fig.~\ref{fig:covfunctions} shows the functional forms of the covariance functions. They have similar shapes: they all return a finite covariance at zero distance which decreases monotonically with distance and tends to zero as the distance becomes large. In the case of interpolating waveform differences this indicates that the errors in the approximate waveform at two nearby points in parameter space are closely related, whereas the errors at two well separated points are nearly independent. The PLE, Cauchy and Mat\'{e}rn function can all be viewed as attempts to generalise the SE with the inclusion of one extra hyperparameter $\eta$, to allow for more flexible GP modelling. All three alternative functions are able to recover the SE in some limiting case, but the Mat\'{e}rn is the most flexible of the three. This can be seen from the discussion of the MS differentiability of GPs given at the beginning of this section. 

The SE covariance function is infinitely differentiable at $\tau=0$, and so the corresponding GP is infinitely MS differentiable. The PLE function is infinitely differentiable at $\tau=0$ for the SE case when $\eta=2$, but for all other cases it is not at all MS differentiable. In contrast, the Cauchy function is infinitely differentiable at $\tau=0$ for all choices of the hyperparameter $\eta$. The Mat\'{e}rn function, by contrast, has a variable level of differentiability at $\tau=0$, controlled via the hyperparameter $\eta$ \cite{Stein1999}. The GP corresponding to the Mat\'{e}rn covariance function in Eq.~\eqref{eq:covMatern} is $n_{d}$-times MS differentiable if and only if $\eta>n_{d}$. This ability to adjust the differentiability allows the same covariance function to successfully model a wide variety of data. In the process of maximising the hyperlikelihood for the training set over hyperparameter $\eta$, the GP \emph{learns} the (non)smoothness properties favoured by the data, and the GPR returns a correspondingly (non)smooth function. 

\subsection{The inclusion of noise $\sigma_{n}$}\label{subsec:covnoise}

Even the most accurate waveform models $h(\vec{\lambda})$ still contain some error with respect to the unknown true physical signal $h'(\vec{\lambda})$. This could be because the waveform model does not include all of the physics or because it is calculated using a method with finite accuracy. We can account for the error in our training set points by adding a noise variance term $\sigma_f^2\sigma_{n,\,i}^2$ in the covariance function,
\begin{equation}\label{eq:noise}
k(\vec{\lambda}_{i},\vec{\lambda}_{j})\to k(\vec{\lambda}_{i},\vec{\lambda}_{j})+\sigma_f^2\sigma_{n,\,i}^2\delta_{ij}\,,
\end{equation}
which alters the covariance matrix in Eq.~\eqref{eq:covmatrix} correspondingly, but not the expressions in Eq.~\eqref{eq:vecandscalar}. Here $\sigma_{n,\,i}$ is the \emph{fractional} error $\|h-h'\|/\|\delta h\|$ in each training set point, where the norm is taken with respect to the inner product in Eq.~\eqref{eq:innerprod} and $\delta h=H-h$. This ensures that $\sigma_f^2$ is still an overall scale for the covariance function.

We do not maximise the hyperlikelihood over $\sigma_{n,\,i}^2$; this is because $\sigma_{n,\,i}$ is related to $\|h-h'\|$, which cannot be learnt from a training set containing the differences $\delta h$. The noise error is instead fixed at some overall error estimate for the accurate model, which is a conservative approach. We consider the simple case $\sigma_{n,\,i}=\sigma_n$ in this paper; however, it is not necessary for all training set points to have the same error, as a training set might comprise different families of waveform models (e.g., a mix of different variants of (S)EOBNR or IMRPhenom waveforms, or NR waveforms with different numerical resolutions).

If the overall noise error is $\sigma_f\sigma_n$, then the GPR uncertainty at a training set point $\sigma(\vec{\lambda}_i)$ satisfies,
\begin{equation}\label{eq:boundonerror}
\sigma(\vec{\lambda}_i)\leq\sigma_f\sigma_n,\quad\forall i\in\{1,2,\ldots,N\}.
\end{equation}
This is because the different points in the training set are assumed to come from a correlated GP, and so nearby measurements also act to decrease the error.

There is also a more practical motivation for the inclusion of noise. Inversion of the covariance matrix in Eq.~\eqref{eq:covmatrix} can pose issues of numerical stability for large training sets. In general, as the number of points in the training set increases, the determinant of the covariance matrix decreases rapidly towards zero, such that the matrix is nearly singular (and hence the matrix is difficult to invert). The solution to this is to add a small fixed noise $\sigma_n^2=J\ll1$, or \emph{jitter}, to each training set point as per Eq.~\eqref{eq:noise}. The eigenvalues of the new covariance matrix are then (approximately) the eigenvalues of the original matrix plus $J$. This prevents the determinant, the product of the eigenvalues, from becoming vanishingly small and dramatically improves the stability of the inversion. In effect, we are no longer requiring our interpolant to pass through every training set point; instead, we only ask it to pass close to each point (with the proximity determined by the value of $J$).

\subsection{Compact support and sparseness}\label{subsec:WENDLANDcovfuncs}

All of the covariance functions considered up until this point have been strictly positive;
\begin{equation}
k(\tau)>0 \quad \forall \; \tau \in [ 0,\infty ) \,.
\end{equation}
When evaluating the covariance matrix for the training set $K_{ij}$ this leads to a matrix where all entries are positive; i.e.\ a dense matrix. When performing the GPR it is necessary to maximise the hyperlikelihood for the training set with respect to the hyperparameters. This process involves inverting the dense matrix $K_{ij}$ at each iteration of the optimisation algorithm. Although this procedure is carried out offline, it still can become prohibitive for large training sets. A related problem, as pointed out in Sec.~\ref{subsec:covnoise} is that for large training sets the determinant of the covariance matrix is typically small which also contributes to making the covariance matrix hard to invert.

One potential way around these issues is to consider a covariance function with compact support,
\begin{equation}
\begin{split}
k(\tau) &> 0 \quad \tau \in [ 0,T ]\,,\\
k(\tau) &= 0 \quad \forall \; \tau \in (T,\infty ) \,,
\end{split} \label{eq:compactsupport}
\end{equation}
where $T$ is some threshold distance beyond which we assume that the waveform differences become uncorrelated. This leads to a sparse, band-diagonal covariance matrix, which is much easier to invert. Care must be taken when specifying the covariance function to ensure that the function is still positive definite (which is required of a GP): if the SE covariance function is truncated, then the matrix formed from the new covariance function is not guaranteed to be positive definite.

\begin{figure}
 \centering
 \includegraphics[trim=0cm 0cm 0cm 0cm, width=0.45\textwidth]{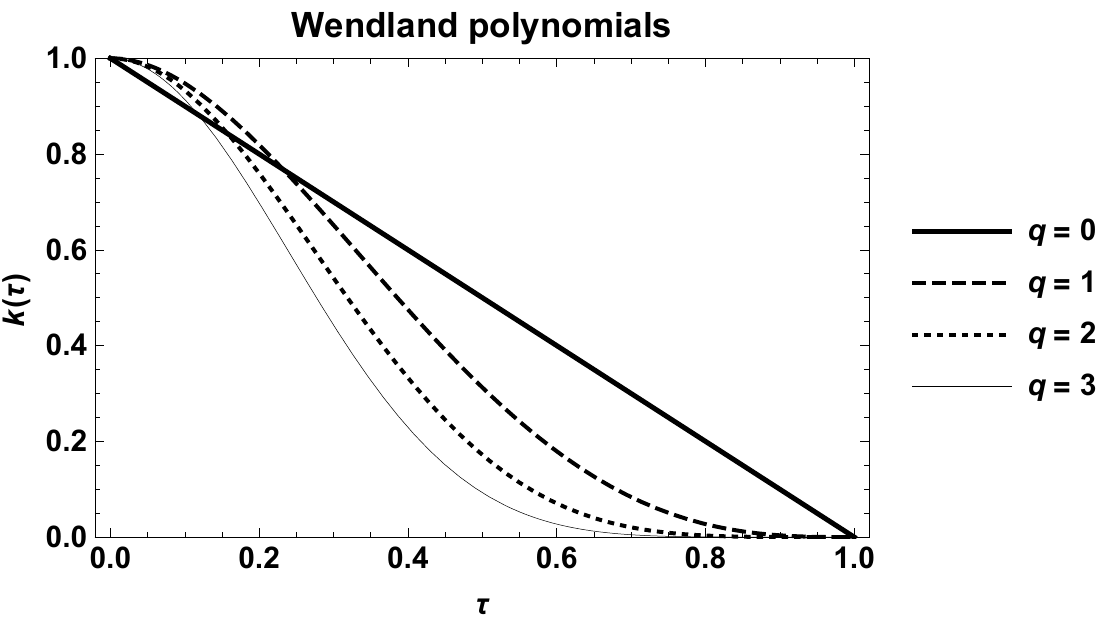}
 \caption{Plots of the first few Wendland polynomial covariance functions. All these functions have compact support, $k(\tau)=0$ for $\tau>1$. As the value of $q$ increases the functions become smoother near the origin.}
 \label{fig:WENDLANDcovfunctions}
\end{figure}

Nevertheless, it is possible to construct covariance functions which have the requisite properties and satisfy the compact support condition in Eq.~\eqref{eq:compactsupport}. These are typically based on polynomials. We consider a series of polynomials proposed by \cite{Wendland}, which we will refer to as the \emph{Wendland} polynomials. These have the property that they are positive definite in $\mathbb{R}^{D}$ and are $2q$-time differentiable at the origin. Therefore the discrete parameter $q$ is in some sense analogous to the $\eta$ hyperparameter of the Mat\'ern covariance function in that it controls the smoothness of the GP. Defining $\beta$ to be 
\begin{equation}
\beta=\left \lfloor{\frac{D}{2}}\right \rfloor +q+1  \,
\end{equation}
and using $\Theta (x)$ to denote the Heaviside step function, the first few Wendland polynomials $k_{D,\,q}(\tau)$ are given by,
\begin{eqnarray}
k_{D,\,0}(\tau) &=& \sigma_{f}^{2}\Theta (1 - \tau)(1 - \tau)^{\beta} \,,\\
k_{D,\,1}(\tau) &=& \sigma_{f}^{2}\Theta (1 - \tau)(1 - \tau)^{\beta+1}\left[1+\left(\beta+1\right)\tau\right] \,,\\
k_{D,\,2}(\tau) &=& \frac{\sigma_{f}^{2}}{3}\Theta (1 - \tau)(1 - \tau)^{\beta+2}\left[\right.\vphantom{\left(\beta^2\right)} 3 + \left(3\beta+6\right)\tau\nonumber\\
&& \left.+\left(\beta^{2}+4\beta+3\right)\tau^{2}\right] \,,\\
k_{D,\,3}(\tau) &=& \frac{\sigma_{f}^{2}}{15}\Theta (1 - \tau)(1 - \tau)^{\beta+3}\left[\vphantom{\left(\beta^2\right)} 15 + \left(15\beta+45\right)\tau \right.\nonumber\\
&& \left. + \left(6\beta^{2}+36\beta+45\right)\tau^{2}\right.\nonumber\\
&& \left. + \left(\beta^{3}+9\beta^{2}+23\beta+15\right)\tau^{3} \right] \,.
\end{eqnarray}
These are plotted in Fig.~\ref{fig:WENDLANDcovfunctions}. Other types of covariance function with compact support have also been proposed and explored in the literature (e.g., \cite{gneiting2002compactly,melkumyan2009sparse,liang2015class}), but we do not consider them in this paper.

\section{Properties of the method}\label{sec:props}

In this section proofs of several useful features of the marginalised likelihood are presented. In Sec.~\ref{sec:linear} we derive the PE error in a linearised formalism, recovering results of \cite{PhysRevD.76.104018} as well as new results for our marginalised likelihood; in Sec.~\ref{sec:infSNR} we use these results to show that the marginalised likelihood should not exclude the true parameter values even at large SNR, and in Sec.~\ref{sec:across-param}, we derive other limits of the marginalised likelihood at specific points in parameter space.

\subsection{The error at linear order}\label{sec:linear}

A more detailed understanding of the theoretical error problem, and the solution offered by the marginalised likelihood can be gained by examining the behaviour of the likelihoods in the vicinity of a maximum. 

The \emph{exact likelihood}, from Eq.~\eqref{eq:Lexact}, is given by
\begin{equation}
L'(\vec{\lambda})\propto \exp\left(-\frac{1}{2}\left\| s-h(\vec{\lambda})\right\|^2\right)\, ,
\end{equation}
and has a maximum at the best fit parameters, $\vec{\lambda}_{\mathrm{bf}}$, which satisfy the equation
\begin{equation}
\label{eq:star} \left\langle \partial_{a}h(\vec{\lambda}_{\mathrm{bf}})\middle|s-h(\vec{\lambda}_{\mathrm{bf}})\right\rangle=0 \, .
\end{equation}
The measured data consist of noise and the physical signal with the true parameters, $\vec{\lambda}_{\mathrm{tr}}$, that is $s=n+h(\vec{\lambda}_{\mathrm{tr}})$. Therefore Eq.~\eqref{eq:star} becomes
\begin{equation}
\left\langle \partial_{a}h(\vec{\lambda}_{\mathrm{bf}})\middle|n+h(\vec{\lambda}_{\mathrm{tr}})-h(\vec{\lambda}_{\mathrm{bf}})\right\rangle=0 \, .
\end{equation}
Expanding the difference in the signals to leading order in $\Delta\vec{\lambda}=\vec{\lambda}_{\mathrm{bf}}-\vec{\lambda}_{\mathrm{tr}}$ gives
\begin{equation}
\left\langle \partial_{a}h(\vec{\lambda}_{\mathrm{bf}})\middle|n-\Delta\vec{\lambda}^{b}\partial_{b}h(\vec{\lambda}_{\mathrm{bf}})\right\rangle = 0 \, ,
\end{equation}
whence
\begin{equation}
\Delta\vec{\lambda}^{a} = \left(\Sigma^{-1}\right)^{ab}\left\langle n\middle|\partial_{b}h(\vec{\lambda}_{\mathrm{bf}})\right\rangle \, ,
\label{eq:deltalambdaexactL}
\end{equation}
where $\Sigma_{ab}=\langle \partial_{a}h(\vec{\lambda}_{\mathrm{bf}})|\partial_{b}h(\vec{\lambda}_{\mathrm{bf}})\rangle$. Therefore, at leading order, the shift between the best fit and true parameters for the exact likelihood consists of one term proportional to $n$; we call this the noise error. The matrix $\Sigma_{ab}$ is the usual Fisher information matrix (FIM) which characterises the random errors at leading order \cite{2008PhRvD..77d2001V}.

The \emph{approximate likelihood}, from Eq.~\eqref{eq:Lapprox}, is given by
\begin{equation}
L(\vec{\lambda})\propto \exp\left(-\frac{1}{2}\left\| s-H(\vec{\lambda})\right\|^2\right)\, ,
\end{equation}
and has a maximum at the best fit parameters which satisfy the equation
\begin{equation}
\left\langle \partial_{a}H(\vec{\lambda}_{\mathrm{bf}})\middle|s-H(\vec{\lambda}_{\mathrm{bf}})\right\rangle=0 \, . \label{eq:approxtemplin}
\end{equation}
Using ${s=n+h(\vec{\lambda}_\mathrm{tr})}$ in Eq.~\eqref{eq:approxtemplin} and expanding to leading order in $\Delta\vec{\lambda}$ gives
\begin{equation}
\left\langle \partial_{a}H(\vec{\lambda}_{\mathrm{bf}})  \middle|n-\delta h(\vec{\lambda}_{\mathrm{tr}})-\Delta\vec{\lambda}^{b}\partial_{b}H(\vec{\lambda}_{\mathrm{bf}})\right\rangle = 0\,,
\end{equation}
thus
\begin{eqnarray}
\Delta\vec{\lambda}^{a} &=& \left(\Gamma^{-1}\right)^{ab}\left\langle n\middle|\partial_{b}H(\vec{\lambda}_{\mathrm{bf}})\right\rangle \nonumber\\
 & & -\left(\Gamma^{-1}\right)^{ab}\left\langle \delta h (\vec{\lambda}_{\mathrm{tr}})\middle|\partial_{b}H(\vec{\lambda}_{\mathrm{bf}})\right\rangle \, ,\label{eq:deltalambdaapproxL}
\end{eqnarray}
where ${\Gamma_{ab}=\langle \partial_{a}H(\vec{\lambda}_{\mathrm{bf}})|\partial_{b}H(\vec{\lambda}_{\mathrm{bf}})\rangle}$. Therefore, at leading order the shift between the best fit and true parameters for the approximate likelihood consists of two terms: the noise error as before (except with the FIM evaluated with the approximate model) and what we call the model error, 
\begin{equation}
\Delta_{\mathrm{model}}\vec{\lambda}^{a} = -\left(\Gamma^{-1}\right)^{ab}\left\langle \delta h (\vec{\lambda}_{\mathrm{tr}})\middle|\partial_{b}H(\vec{\lambda}_{\mathrm{bf}})\right\rangle\,.
\end{equation}
The model error is independent of the noise realisation, and hence represents a systematic error in the PE associated with using inaccurate models.

The above treatment of the exact and approximate likelihoods is a brief summary of part of the analysis done by \cite{PhysRevD.76.104018}. We now apply the same type of analysis to the new marginalised likelihood to see how this reduces or removes the model error.

The \emph{marginalised likelihood} is given in Eq.~\eqref{eq:Lcal}. From Eq.~\eqref{eq:GPRsigma2} it can be seen that the interpolated waveform difference $\mu(\vec{\lambda})$ is a linear combination of $\delta h(\vec{\lambda}_{i})$ from the training set. We will assume, for this calculation only, that the waveform difference is also a small quantity in the sense that $\|\delta h\|\ll \|h\|$ with the norm from Eq.~\eqref{eq:norm}. Therefore, $\mu={\mathcal{O}}(\delta h)$ and $\sigma_{f}={\mathcal{O}}(\delta h)$. We shall keep contributions up to ${\mathcal{O}}(\delta h)$.

Under the twin assumptions that $\Delta \vec{\lambda}$ and $\|\delta h\|$ are small, the marginalised likelihood is approximately given by
\begin{equation}
{\mathcal{L}}(\vec{\lambda})\approx \exp\left(-\frac{1}{2}\left\| s-H(\vec{\lambda})+\mu(\vec{\lambda})\right\|^2\right)\, .
\end{equation}
This has a maximum at the best fit parameters $\vec{\lambda}_{\mathrm{bf}}$ which satisfy the equation
\begin{equation}
\left\langle \partial_{a}\left(\mu(\vec{\lambda}_{\mathrm{bf}})-H(\vec{\lambda}_{\mathrm{bf}})\right)\middle|s-H(\vec{\lambda}_{\mathrm{bf}})+\mu(\vec{\lambda}_{\mathrm{bf}})\right\rangle=0 \, .
\end{equation}
Using ${s=n+h(\vec{\lambda}_{\mathrm{tr}})}$, and expanding to leading order in $\Delta\vec{\lambda}$ and $\delta h$ gives
\begin{widetext}
\begin{eqnarray}
\left\langle -\partial_{a}\left(\mu(\vec{\lambda}_{\mathrm{bf}})-H(\vec{\lambda}_{\mathrm{bf}})\right)\middle|n+h(\vec{\lambda}_{\mathrm{tr}})-H(\vec{\lambda}_{\mathrm{bf}})+\mu(\vec{\lambda}_{\mathrm{bf}})\right\rangle &=& 0 \,, \\ 
\left\langle -\partial_{a}\left(\mu(\vec{\lambda}_{\mathrm{bf}})-H(\vec{\lambda}_{\mathrm{bf}})\right)\middle|n-\delta h(\vec{\lambda}_{\mathrm{tr}})-\Delta\vec{\lambda}^{b}\partial_{b}H(\vec{\lambda}_{\mathrm{bf}})+\mu(\vec{\lambda}_{\mathrm{bf}})\right\rangle &=& 0 \,.
\end{eqnarray}
\end{widetext}
This expression can be rearranged to find $\Delta\vec{\lambda}$, dropping all terms second order in small quantities,
\begin{eqnarray}
\Delta\vec{\lambda}^{a} &=& \left(\Gamma^{-1}\right)^{ab}\left\langle n\middle|\partial_{b}\left(H(\vec{\lambda}_{\mathrm{bf}})-\mu(\vec{\lambda}_{\mathrm{bf}})\right)\right\rangle \nonumber\\
&& -\left(\Gamma^{-1}\right)^{ab}\left\langle \delta h (\vec{\lambda}_{\mathrm{tr}})\middle|\partial_{b}H(\vec{\lambda}_{\mathrm{bf}})\right\rangle \nonumber\\
&& +\left(\Gamma^{-1}\right)^{ab}\left\langle \mu(\vec{\lambda}_{\mathrm{bf}})\middle|\partial_{b}H(\vec{\lambda}_{\mathrm{bf}})\right\rangle \, . \label{eq:deltalambdamargL}
\end{eqnarray}
Therefore, at leading order, the shift between the best fit and true parameters for the marginalised likelihood consists of three terms: the noise and model errors from before, and a new shift arising from the marginalisation, $\Delta_{\mathrm{marg}}\vec{\lambda}^{a}~=~\left(\Gamma^{-1}\right)^{ab}  \langle \mu(\vec{\lambda}_{\mathrm{bf}})|\partial_{b}H(\vec{\lambda}_{\mathrm{bf}})\rangle$. The expression for the model and marginalisation errors are similar and appear with opposite signs (as would be hoped since the maginalised likelihood was designed to remove the model error) so the remaining model error is proportional to $\delta h(\vec{\lambda}_{\mathrm{bf}})-\mu(\vec{\lambda}_{\mathrm{bf}})$ (integrated inside the inner product).

If the training set is dense in the region of the peak, and the hyperparameters have been correctly estimated, it is reasonable to assume that the GPR interpolant of the waveform difference performs well, and we have ${\langle \delta h(\vec{\lambda})-\mu(\vec{\lambda})|\cdot\rangle \approx 0}$. Under these conditions the marginalised likelihood removes the systematic model error from the parameter estimates. In reality the interpolation is not perfect, and the method is limited by the available information in the training set, so that a residual model error proportional to $\langle\delta h(\vec{\lambda}_{\mathrm{bf}})-\mu(\vec{\lambda}_{\mathrm{bf}})|\cdot\rangle$ remains.

\subsection{The limit of large SNR}\label{sec:infSNR}

As first pointed out by \cite{PhysRevD.76.104018}, the systematic error associated with the inaccurate model used in the approximate likelihood is independent of the SNR, whereas the random error associated with the noise realisation decreases with increasing SNR. Therefore, there exists a critical SNR for the approximate likelihood above which the systematic model error dominates the random noise error. If the approximate likelihood is used to infer the parameters of a source with an SNR close to or above this critical value then the inferred parameters are significantly and systematically biased. In this section we examine the behaviour of all three likelihood functions for large SNR and show that the marginalised likelihood does \emph{not} suffer from this problem even in the limit of infinite SNR. Therefore, parameter estimates obtained using the marginalised likelihood can always be trusted.

In this section in order to ease the process of taking the limit of large SNR all waveforms are understood to be normalised such that $\|h(\vec{\lambda})\|=1$, and the amplitude is taken out as a prefactor, so the full signal is $Ah(\vec{\lambda})$. In addition we will assume for simplicity that the measured value of $A$ is equal to the true value for the signal; this has no effect our final result.

The \emph{exact likelihood} Eq.~\eqref{eq:Lexact} is given by
\begin{equation}
L'(\vec{\lambda})\propto \exp\left(-\frac{1}{2}\left\| s-Ah(\vec{\lambda})\right\|^2\right)\, .
\end{equation}
The measured data is given by $s=n+Ah(\vec{\lambda}_{\mathrm{tr}})$, and the exact likelihood is peaked at $\vec{\lambda}_{\mathrm{bf}}=\vec{\lambda}_{\mathrm{tr}}+\Delta\vec{\lambda}$, where (see Eq.~\eqref{eq:deltalambdaexactL})
\begin{equation}
\Delta\vec{\lambda}^{a}=\frac{1}{A}\left(\Sigma^{-1}\right)^{ab}\left\langle n\middle|\partial_{b}h(\vec{\lambda}_{\mathrm{bf}})\right\rangle\,.
\end{equation}
In this section, the FIM $\Sigma_{ab}$ is defined in terms of the normalised waveforms, i.e.\ $\Sigma_{ab}$ is independent of $A$; this is done so that all of the dependence on $A$ remains explicit. The exact likelihood evaluated on the true parameters is given by
\begin{equation}
L'(\vec{\lambda}_{\mathrm{tr}}) \propto \exp\left(-\frac{1}{2}\left\| n \right\|^2\right)\, .
\end{equation}
The exact likelihood evaluated on the best-fit parameters is given by
\begin{equation}
L'(\vec{\lambda}_{\mathrm{bf}}) \propto \exp\left[-\frac{1}{2}\left\| n+A\left(h(\vec{\lambda}_\mathrm{tr})-h(\vec{\lambda}_\mathrm{bf})\right) \right\|^2\right]\, .
\end{equation}
The ratio of these two likelihood values is denoted $R_{\mathrm{exact}}=L'(\vec{\lambda}_{\mathrm{tr}})/L'(\vec{\lambda}_{\mathrm{bf}})$. Expanding the difference $h(\vec{\lambda}_\mathrm{tr})-h(\vec{\lambda}_\mathrm{bf})$ in the above equation to leading order in $\Delta\vec{\lambda}$ gives
\begin{equation}
\ln R_{\mathrm{exact}} =-\frac{1}{2}\left(\Sigma^{-1}\right)^{ab}\left\langle n\middle| \partial_{a}h(\vec{\lambda}_{\mathrm{bf}})\right\rangle\left\langle n\middle| \partial_{b}h(\vec{\lambda}_{\mathrm{bf}})\right\rangle\, .
\label{eq:Rexact}
\end{equation}
The quantity $R_{\mathrm{exact}}$ is the factor by which the likelihood of the true parameters is suppressed with respect to the peak likelihood. From Eq.~\eqref{eq:Rexact} it can be seen that this factor is a random variable dependent on the noise realisation $n$; the expectation of this random variable is given by \cite{1994PhRvD..49.2658C}
\begin{equation}
\overline{\ln R_{\mathrm{exact}}} = -\frac{1}{2}\, .
\label{eq:meanRexact}
\end{equation}
Both Eqs.~\eqref{eq:meanRexact} and Eq.~\eqref{eq:Rexact} are independent of the signal amplitude $A$, and hence are unchanged by taking the limit of large SNR, $A\rightarrow\infty$. Therefore (as one might have expected) the exact likelihood evaluated at $\vec{\lambda}_{\mathrm{tr}}$ remains finite in this limit and the true parameters are never completely excluded from the posterior at any value of the SNR.

The \emph{approximate likelihood} Eq.~\eqref{eq:Lapprox} is given by
\begin{equation}
L(\vec{\lambda})\propto \exp\left(-\frac{1}{2}\left\| s-AH(\vec{\lambda})\right\|^2\right)\, ,
\end{equation}
The approximate likelihood is peaked at $\vec{\lambda}_{\mathrm{bf}}=\vec{\lambda}_{\mathrm{tr}}+\Delta\vec{\lambda}$, where (see Eq.~\eqref{eq:deltalambdaapproxL})
\begin{eqnarray}
\Delta\vec{\lambda}^{a}&=&\frac{1}{A}\left(\Gamma^{-1}\right)^{ab}\left\langle n\middle|\partial_{b}H(\vec{\lambda}_{\mathrm{bf}})\right\rangle\nonumber\\
&&-\left(\Gamma^{-1}\right)^{ab}\left\langle \delta h (\vec{\lambda}_{\mathrm{tr}})\middle|\partial_{b}H(\vec{\lambda}_{\mathrm{bf}})\right\rangle\, .
\label{eq:deltalambdaapprox123}
\end{eqnarray}
The FIM $\Gamma_{ab}$ is also here defined to be independent of $A$. The approximate likelihood evaluated on the true parameters is given by
\begin{eqnarray}
L(\vec{\lambda}_{\mathrm{tr}}) &\propto &\exp\left[-\frac{1}{2}\left\|  n+A\left(h(\vec{\lambda}_\mathrm{tr})-H(\vec{\lambda}_\mathrm{tr})\right) \right\|^2\right]\nonumber\\
&\propto&\exp\left(-\frac{1}{2}\left\|  n-A\delta h(\vec{\lambda}_\mathrm{tr})\right\|^2\right)  \, .
\label{eqLapproxtr}
\end{eqnarray}
The approximate likelihood evaluated on the best fit parameters is given by
\begin{widetext}
\begin{eqnarray} 
L(\vec{\lambda}_{\mathrm{bf}}) &\propto& \exp\left[-\frac{1}{2}\left\| n-A\left(h(\vec{\lambda}_\mathrm{tr})-H(\vec{\lambda}_\mathrm{bf})\right)\right\|^2\right] \nonumber \\
&\propto& \exp\left[-\frac{1}{2}\left\| n+A\left(\delta h(\vec{\lambda}_{\mathrm{tr}})-\Delta\vec{\lambda}^{a}\partial_{a}H(\vec{\lambda}_{\mathrm{bf}})\right)\right\|^2\right] \,,
\label{eqLapproxbf}
\end{eqnarray}
\end{widetext}
where, as before, the waveform difference has been expanded to leading order in $\Delta\vec{\lambda}$. The ratio of the two likelihood $R_{\mathrm{approx}}=L(\vec{\lambda}_{\mathrm{tr}})/L(\vec{\lambda}_{\mathrm{bf}})$ can be evaluated from Eq.~\eqref{eqLapproxtr} and Eq.~\eqref{eqLapproxbf}, and taking the limit of large SNR gives
\begin{eqnarray}
\lim_{A\rightarrow\infty}\ln R_{\mathrm{approx}} &=& -\frac{A^{2}}{2}\left(\Gamma^{-1}\right)^{ab} \left\langle \delta h(\vec{\lambda}_{\mathrm{tr}})\middle|\partial_{a}H(\vec{\lambda}_{\mathrm{bf}})\right\rangle \nonumber\\
& & \times \left\langle \delta h(\vec{\lambda}_{\mathrm{tr}})\middle|\partial_{b}H(\vec{\lambda}_{\mathrm{bf}})\right\rangle\,.
\end{eqnarray}
Unlike $R_{\mathrm{exact}}$, this ratio does not depend on $n$. This is because in the limit of large SNR, only the terms from the exponents of Eq.~\eqref{eqLapproxtr} and Eq.~\eqref{eqLapproxbf} proportional to $A^{2}$ contribute, and the noise-dependent terms are all proportional to $A$. Also unlike $R_{\mathrm{exact}}$, this ratio does depend on the amplitude and $R_{\mathrm{approx}}\rightarrow 0$ in the limit of large SNR. Therefore, as anticipated above, the approximate likelihood excludes the true source parameters with complete certainty in the limit of large SNR (unless $\langle\delta h(\vec{\lambda}_{\mathrm{tr}})|\cdot\rangle =0$, in which case the exact likelihood is recovered).

The \emph{marginalised likelihood} Eq.~\eqref{eq:Lcal} is given by
\begin{equation}
\label{eq:tempreflcal}
{\mathcal{L}}(\vec{\lambda})\propto \exp\left(-\frac{1}{2}\frac{\left\| s-AH(\vec{\lambda})+A\mu(\vec{\lambda})\right\|^2}{1+A^{2}\sigma^{2}(\vec{\lambda})}\right)\, .
\end{equation}
The marginalised likelihood is peaked at $\vec{\lambda}_{\mathrm{bf}}=\vec{\lambda}_{\mathrm{tr}}+\Delta\vec{\lambda}$, where, by comparison with Eq.~\eqref{eq:deltalambdamargL},
\begin{eqnarray}\Delta\vec{\lambda}^{a}&=&\frac{1}{A}\left(\Gamma^{-1}\right)^{ab}\left\langle n\middle|\partial_{b}\left(H(\vec{\lambda}_{\mathrm{bf}})-\mu(\vec{\lambda}_{\mathrm{bf}})\right)\right\rangle\nonumber\\
&&-\left(\Gamma^{-1}\right)^{ab}\left\langle \delta h (\vec{\lambda}_{\mathrm{tr}})\middle|\partial_{b}H(\vec{\lambda}_{\mathrm{bf}})\right\rangle\nonumber\\
&&+\left(\Gamma^{-1}\right)^{ab}\left\langle \mu(\vec{\lambda}_{\mathrm{bf}})\middle|\partial_{b}H(\vec{\lambda}_{\mathrm{bf}})\right\rangle\, .\label{eq:deltalambdamarg123}\end{eqnarray}
The values of the marginalised likelihood evaluated on the true and best-fit parameters are given by Eq.~\eqref{eqLmargtr} and Eq.~\eqref{eqLmargbf}, and the ratio of these two likelihoods is denoted $R_{\mathrm{marg}}$,
\begin{widetext}
\begin{eqnarray}
\label{eqLmargtr}
{\mathcal{L}}(\vec{\lambda}_{\mathrm{tr}}) &\propto& \exp\left(-\frac{1}{2}\frac{\left\| n-A\delta h(\vec{\lambda}_{\mathrm{tr}})+A\mu(\vec{\lambda}_{\mathrm{tr}})\right\|^2}{1+A^{2}\sigma^{2}(\vec{\lambda}_{\mathrm{tr}})}\right)\,; \\
%
{\mathcal{L}}(\vec{\lambda}_{\mathrm{bf}}) &\propto& \exp\left(-\frac{1}{2}\frac{\left\| n-A\delta h(\vec{\lambda}_{\mathrm{tr}})-A\Delta\vec{\lambda}^{a}\partial_{a}H(\vec{\lambda}_{\mathrm{bf}})+A\mu(\vec{\lambda}_{\mathrm{bf}})\right\|^2}{1+A^{2}\sigma^{2}(\vec{\lambda}_{\mathrm{bf}})}\right)\,;
\label{eqLmargbf} \\
\lim_{A\rightarrow\infty}\ln R_{\mathrm{marg}} &=& -\frac{1}{2\sigma^{2}(\vec{\lambda}_{\mathrm{bf}})} \left[
\left(\Gamma^{-1}\right)^{ab}\left\langle \delta h (\vec{\lambda}_{\mathrm{tr}})-\mu(\vec{\lambda}_{\mathrm{bf}})\middle|\partial_{a}H(\vec{\lambda}_{\mathrm{bf}})\right\rangle\left\langle \delta h (\vec{\lambda}_{\mathrm{tr}})-\mu(\vec{\lambda}_{\mathrm{bf}})\middle|\partial_{b}H(\vec{\lambda}_{\mathrm{bf}})\right\rangle\right.\nonumber\\
&&\left. -\left\| \mu (\vec{\lambda}_{\mathrm{bf}})\right\|^2 + \left\| \mu (\vec{\lambda}_{\mathrm{tr}})\right\|^2 +2\left\langle \delta h (\vec{\lambda}_{\mathrm{tr}})\middle|\mu (\vec{\lambda}_{\mathrm{tr}})-\mu (\vec{\lambda}_{\mathrm{bf}})\right\rangle \right] \label{eq:limmargratio1}\\
 &\approx& -\frac{1}{2\sigma^{2}(\vec{\lambda}_{\mathrm{bf}})}\left(\Gamma^{-1}\right)^{ab}\left\langle \delta h (\vec{\lambda}_{\mathrm{tr}})-\mu(\vec{\lambda}_{\mathrm{bf}})\middle|\partial_{a}H(\vec{\lambda}_{\mathrm{bf}})\right\rangle\left\langle \delta h (\vec{\lambda}_{\mathrm{tr}})-\mu(\vec{\lambda}_{\mathrm{bf}})\middle|\partial_{b}H(\vec{\lambda}_{\mathrm{bf}})\right\rangle \,.
\label{eq:limmargratio2}
\end{eqnarray}
\end{widetext}
The approximation made in going from Eq.~\eqref{eq:limmargratio1} to Eq.~\eqref{eq:limmargratio2} involves dropping terms which are products of small quantities. Because the FIM is a symmetric, positive-definite matrix, the numerator in Eq.~\eqref{eq:limmargratio2} is a negative number, and hence $R_{\mathrm{marg}}<1$ as required to ensure $\vec{\lambda}_{\mathrm{bf}}$ is the peak of the likelihood.

As was the case with $R_{\mathrm{approx}}$, this expression for $R_{\mathrm{marg}}$ does not depend on the noise. However, unlike $R_{\mathrm{approx}}$ the expression for $R_{\mathrm{marg}}$ also does not depend on the amplitude $A$. Therefore, in the limit that the SNR becomes large $R_{\mathrm{marg}}$ tends to a constant value which depends quadratically on $\langle\delta h (\vec{\lambda}_{\mathrm{tr}})-\mu(\vec{\lambda}_{\mathrm{bf}})|\cdot\rangle$. As the SNR increases, the true parameters are not excluded from the marginalised likelihood, instead the likelihood distribution tends to a constant distribution (i.e.\ no dependence on $n$), and the ratio by which the true parameters are disfavoured compared to the best fit parameters is set by the ability of the GPR to recover the true waveform difference.

Intuitively, the reason the marginalised likelihood is able to achieve this useful behaviour, even if the true waveform difference is not perfectly recovered by the GPR interpolation (i.e.\ $\langle\delta h (\vec{\lambda}_{\mathrm{tr}}) - \mu(\vec{\lambda}_{\mathrm{bf}})|\cdot\rangle \neq 0$), is due to the way the hyperparameters in the covariance function are chosen. The hyperparameters were fixed to their optimum values by maximising the hyperlikelihood for the training set (as described in Sec.~\ref{sec:method}). During this process the overall scale hyperparameter $\sigma_{f}$ gains a dependence on the amplitude proportional to $A^{2}$. Hence the GPR uncertainty $\sigma^{2}(\vec{\lambda})$ is also proportional to $A^{2}$. As can be seen from Eq.~\eqref{eq:tempreflcal}, in the limit of large SNR the amplitude dependence cancels in the exponential and the marginalised likelihood tends to a constant distribution. Therefore, the marginalised likelihood never excludes the true source parameters from the final posterior with complete certainty.

\subsection{Limits of the marginalised likelihood across parameter space}\label{sec:across-param}

In this section we examine the behaviour of the marginalised likelihood in the limit of being far from any training points and being at a training point.

First we examine the behaviour of the marginalised likelihood in the former case, at a large distance ($\tau^{2}\gg 1$) from any of the points in the training set. From Eq.~\eqref{eq:GPRsigma2} it can be seen that well outside of the training set $\mu(\vec{\lambda})\rightarrow 0$ and $\sigma^{2}(\vec{\lambda})\rightarrow \sigma_{f}^{2}$. Therefore, from Eq.~\eqref{eq:finalresult}, the log marginalised likelihood tends to
\begin{equation}
\ln {\mathcal{L}}(\vec{\lambda})\rightarrow \frac{\ln L(\vec{\lambda}) }{1+\sum_{i,j}{\boldsymbol{K}}_{ij}\left\langle \delta h(\vec{\lambda}_{i})\middle| \delta h(\vec{\lambda}_{j})\right\rangle} \, .
\label{eq:outsideD}
\end{equation}
Well outside of the training set the marginalised likelihood $\ln{\mathcal{L}}(\vec{\lambda})$ recovers the standard, approximate likelihood $L(\vec{\lambda})$ up to a constant factor. This constant factor is one plus a linear combination of the overlap integrals of all the waveform differences in the training set. Since the denominator in Eq.~\eqref{eq:outsideD} is always greater than unity (this is ensured by the positive-definite property of the covariance matrix), it broadens any peak in the likelihood outside of the training set. The amount of the broadening is set by the magnitude of the waveform differences in the training set via the overlap matrix $\langle \delta h(\vec{\lambda}_{i})|\delta h(\vec{\lambda}_{j})\rangle$. This is the behaviour that would be expected; in the absence of any accurate waveforms the parameter uncertainties obtained from the approximate waveforms should be multiplied by a constant factor depending upon our level of belief in the accuracy of the approximate waveform model. In turn, our level of belief in the accuracy of the approximate waveform is learnt from the training set in the process of training the GP.

We now consider the behaviour of the marginalised likelihood evaluated at one of the training set points $\vec{\lambda_{\ell}}$. First, consider the case where $\sigma_{n}=0$. In this case, the interpolated waveform difference, from Eq.~\eqref{eq:GPRmeanequ}, at $\vec{\lambda}_{\ell}$ recovers the true waveform difference, and the GPR uncertainty, from Eq.~\eqref{eq:GPRsigma2}, vanishes at $\vec{\lambda}_{\ell}$;
\begin{eqnarray}
\mu(\vec{\lambda}_{\ell})&=& \delta h(\vec{\lambda}_{\ell}) \, ,\label{eq:onDsetpoint1}\\
\sigma^{2}(\vec{\lambda}_{\ell})&=& 0\, .\label{eq:onDsetpoint2}
\end{eqnarray}
Therefore the marginalised likelihood in Eq.~\eqref{eq:finalresult} recovers the exact likelihood with no additional broadening.
\begin{equation}
{\mathcal{L}}(\vec{\lambda}_{\ell})=L'(\vec{\lambda}_{\ell}) \, .
\end{equation}
This is also the behaviour that would be expected; at a point in parameter space where the accurate waveform is known, the accurate likelihood is recovered.

If $\sigma_{n}\neq 0$
, then Eq.~\eqref{eq:onDsetpoint1} and Eq.~\eqref{eq:onDsetpoint2} become
\begin{eqnarray}
\mu(\vec{\lambda}_{\ell})&=& \delta h(\vec{\lambda}_{\ell}) - \sigma_{n}^{2} \sum_i k(\vec{\lambda}_{i},\vec{\lambda}_{\ell})\delta h(\vec{\lambda}_{i})+{\mathcal{O}}(\sigma_{n}^{4})\, ,\nonumber\\
\sigma^{2}(\vec{\lambda}_{\ell})&=& \simga_{n}^{2} \sum_i k(\vec{\lambda}_{i},\vec{\lambda}_{\ell})k(\vec{\lambda}_{i},\vec{\lambda}_{\ell})+{\mathcal{O}}(\sigma_{n}^{4})\, .
\end{eqnarray}
In this case any peak in the marginalised likelihood will be slightly shifted and broadened relative to the peak in the accurate likelihood by an amount consistent with the uncertainty $\sigma_{n}$ in the accurate waveform model.

\section{Implementation}\label{sec:allres}

In this section we present an illustrative implementation of our GPR approach. As a simple example, we consider estimating a single parameter; a full multidimensional application that would be appropriate for actual GW data analysis will be investigated in future work. We begin in Sec.~\ref{sec:model} by introducing the waveforms we use for this study. In Sec.~\ref{subsec:dset} we describe the placement of the training set points for the GPR; in order to investigate the effect of training set on the GPR interpolant two sets were constructed with different numbers of points and grid spacings. In Sec.~\ref{sec:hyperparams} we present results for maximising the hyperlikelihood to find the optimum hyperparameters, $\vec{\theta}_{\mathrm{op}}$, for the interpolation; this is done for a range of different covariance functions on each of the training sets described in Sec.~\ref{subsec:dset}. In Sec.~\ref{sec:wavematch} we interpolate the waveforms across parameter space for the different training sets and different covariance functions described and compare the interpolated waveforms $H(\vec{\lambda})-\mu(\vec{\lambda})$ to the accurate waveforms $h(\vec{\lambda})$. In Sec.~\ref{sec:waveerr} we present results for the GPR uncertainty, $\sigma^{2}(\vec{\lambda})$, for the different training sets and different covariance functions considered. Finally in Sec.~\ref{sec:likelihoodresults} we present results for the marginalised likelihood ${\mathcal{L}}(\vec{\lambda})$, and compare with the results obtained using the approximate likelihood $L(\vec{\lambda})$, and the exact likelihood $L'(\vec{\lambda})$.

\subsection{Model waveforms}\label{sec:model}

In order to implement the GPR, a choice has to be made regarding which waveform models to use. The method uses two waveform approximants; the accurate $h(\vec{\lambda})$ and the approximate $H(\vec{\lambda})$ waveforms. The accurate waveform should be the most accurate available at a computational cost that permits the offline construction of the training set ${\mathcal{D}}$. The criteria for choosing the approximate waveform is less clear, a balance needs to be struck between accuracy and speed. If the model is computationally cheap but not accurate enough the waveform difference, $\delta h(\vec{\lambda})=H(\vec{\lambda})-h(\vec{\lambda})$, will be large and vary on short length scales over parameter space; these are the situations which will cause the GPR to perform worst. On the other hand an accurate model which is too computationally expensive could slow down any PE to such an extent that there ceases to be any benefit in using the marginalised likelihood instead of the accurate likelihood.

We used two waveform models implemented in the LIGO Scientific Collaboration Algorithm Library (LAL).\footnote{\href{http://www.lsc-group.phys.uwm.edu/lal}{http://www.lsc-group.phys.uwm.edu/lal}} As our intention here is to provide a proof of principle, we choose the IMRPhenomC approximant \cite{PhysRevD.82.064016} as the accurate waveform and the widely used TaylorF2 approximant \cite{2001PhRvD..63d4023D,2002PhRvD..66b7502D,PhysRevD.80.084043} as the approximate waveform; both of these models are sufficiently fast to evaluate that we can compute and then compare the three likelihoods (accurate $L'(\vec{\lambda})$, approximate $L(\vec{\lambda})$, and marginalised ${\mathcal{L}}(\vec{\lambda})$) and directly assess the performance of the GPR.

Both of the approximants we have chosen to use here are frequency-domain models, i.e.\ they naturally return the waveform in the Fourier domain $\tilde{h}(f)$.\footnote{In previous work \cite{Letter} the marginalised likelihood has been implemented with time-domain approximants. The method works equally with frequency-domain or time-domain models without the need to transform between them. In the offline stage the waveforms enter only via the overlap matrix $\langle \delta h(\vec{\lambda}_{i})|\delta h(\vec{\lambda}_{j})\rangle$, and in the online stage the waveforms enter only in the linear combination for $\mu(\vec{\lambda})$ in Eq.~\eqref{eq:GPRmeanequ}, which commutes with the operation of taking the Fourier transform.} The IMRPhenomC waveform includes inspiral, merger and ringdown, while the TaylorF2 waveform only includes the inspiral.

We investigate the merger of non-spinning circular binaries. This limits the number of intrinsic parameters describing the system to two, the masses of the two component objects, $\vec{\lambda}=\{m_{1},m_{2}\}$. To further simplify the problem we place training set points only along a one-dimensional subspace, which we choose to be a surface of constant mass ratio, $Q = m_{2}/m_{1}$ (with $m_{1}\geq m_{2}$), parametrised by the value of the chirp mass $\mathcal{M}_\mathrm{c}=(m_{1}m_{2})^{3/5}/(m_{1}+m_{2})^{1/5}$. This keeps the size of the training set small, and hence the computational complexity of the GPR to a minimum. This allows us to instead focus our attention on the novel features of the marginalised likelihood, and explore the effect of changing various features of the method.

\subsection{The training set}\label{subsec:dset}

For simplicity we restrict the range of the coordinates which we search over to reduce the computational complexity. This is again to allow us to focus our attention on the novel features of the method. The training sets cover the chirp mass in the range $\mathcal{M}_\mathrm{c}\in\left[5,5.6\right]\Msun$ and the mass ratio is fixed to the (nearly equal mass) $Q=0.75$. The placement of training set points was done as a regular grid in chirp mass with a step size between points of $\Delta \mathcal{M}_\mathrm{c}$.

The chirp-mass range has been chosen to demonstrate the properties of the method. For lower masses, the signal is dominated by the inspiral where both approximants agree well. Therefore, interpolating these small differences would not be a robust test. At higher masses, where the signal is just merger and ringdown, the two approximants are completely different; we get no useful information from the TaylorF2 waveform and may as well interpolate IMRPhenomC directly. We do not anticipate that in practice we would consider waveform differences as significant as the complete absence of merger and ringdown; hence, this example, although only one-dimensional, should be a rigorous test of what waveform uncertainties can be successfully interpolated. Understanding if this continues to be true for the interpolation of a more intricate waveform difference across a higher-dimensional parameter space, must wait for further studies to be completed.

To allow us to explore the effect that the density of points in the training set has on the marginalised likelihood two different values for $\Delta \mathcal{M}_\mathrm{c}$ were considered. This leads to two different training sets whose total number $n$ of points are different; the properties of these two training sets are summarised in Tab.~\ref{tab:tab}. It is expected that the GPR interpolation, and hence the marginalised likelihood, will perform better when using the denser set ${\mathcal{D}}_{1}$.
\begin{table}
\caption{The properties describing the positions of the template waveforms for each of the three training sets used.\label{tab:tab}}
\begin{ruledtabular}
\begin{tabular}{cccc}
 & $\Delta \mathcal{M}_\mathrm{c}$ & $N$ \\
\hline
${\mathcal{D}}_{0}$ & $ 1.0\times 10^{-2}\Msun$ & $ 60$ \\
${\mathcal{D}}_{1}$ & $ 5.0\times 10^{-3}\Msun$ & $ 120$ \\
\end{tabular}
\end{ruledtabular}
\end{table}

Once the training set points $\{\vec{\lambda}_{i}\}$ were specified, both the approximate $H(\vec{\lambda}_{i})$ and accurate $h(\vec{\lambda}_{i})$ waveforms discussed in Sec.~\ref{sec:model} were evaluated at each point, and the waveform differences $\{\delta h(\vec{\lambda}_{i})\}$ stored for use during the GPR interpolation. The matrix of waveform difference overlaps $M_{ij}=\langle \delta h(\vec{\lambda}_{i})|\delta h(\vec{\lambda}_{j})\rangle$ was also evaluated and stored for use during the hyperlikelihood maximisation procedure.

\subsection{The hyperparameters}\label{sec:hyperparams}

\begin{figure}
 \centering
 \includegraphics[width=0.9\columnwidth]{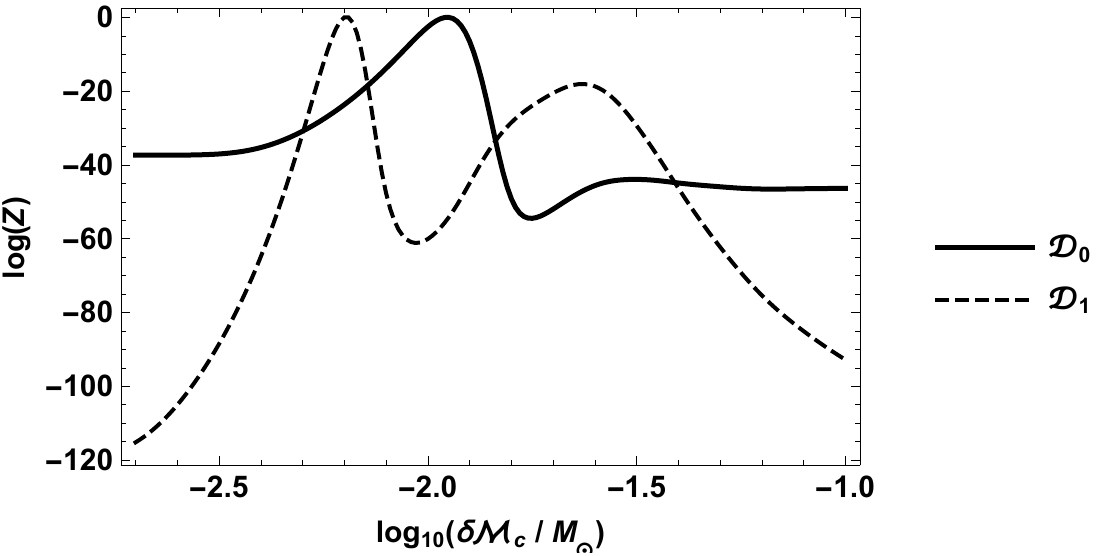}
 \caption{The hyperlikelihood, from Eq.~\eqref{eq:evtrainset}, for the SE covariance function, maximised over the scale hyperparameter $\sigma_{f}$, plotted against the chirp mass length scale $\delta \mathcal{M}_\mathrm{c}$. The hyperlikelihood is shown for both of the training sets (normalised to a peak value of $1$). The denser training set ${\mathcal{D}}_{1}$ was found to favour smaller length scales.}
 \label{fig:Evidence_L}
\end{figure}

Initially the training sets described in Sec.~\ref{subsec:dset} were interpolated using the SE covariance function in Eq.~\eqref{eq:covSE}. This covariance function has just two hyperparameters, $\vec{\theta} = \{ \sigma_{f},g_{\mathcal{M}_\mathrm{c}\mathcal{M}_\mathrm{c}} \}$. The one-dimensional metric $g_{\mathcal{M}_\mathrm{c}\mathcal{M}_\mathrm{c}}$ can be exchanged for a length scale in the chirp mass parameter $\delta \mathcal{M}_\mathrm{c} \equiv 1/\sqrt{g_{\mathcal{M}_\mathrm{c}\mathcal{M}_\mathrm{c}}}$. A fixed noise term with $\sigma_{n}^2=10^{-4}$ was used for all the covariance functions in this section, to make the inverse of the covariance function numerically stable as discussed in Sec.~\ref{subsec:covnoise}. The hyperlikelihood for the training set ${\mathcal{D}}_{0}$ was maximised with respect to these two hyperparameters. The optimum values for the hyperparameters were found to be
\begin{eqnarray}
\sigma_{f} &=& 3.49\times 10^{4} \,,\\ 
\delta \mathcal{M}_\mathrm{c} &=& 1.11 \times 10^{-2} \Msun\,.
\end{eqnarray}
The hyperlikelihood is shown in Fig.~\ref{fig:Evidence_L}. The hyperlikelihood was also maximised for the training set ${\mathcal{D}}_{1}$ using the same SE covariance function and those results are also shown in Fig.~\ref{fig:Evidence_L}. For the denser training set ${\mathcal{D}}_{1}$ the optimum length scale was found to be smaller, $\delta\mathcal{M}_\mathrm{c}=6.31\times 10^{-3}\Msun$. For both training sets, in the limit that the length scale becomes much larger than the total width of the training set ($0.6\Msun$) or much smaller than the grid point spacing ($\Delta \mathcal{M}_\mathrm{c}$), the hyperlikelihood tends to a constant value. This behaviour can be understood by examining the expression for the hyperlikelihood in Eq.~\eqref{eq:evtrainset}.

\begin{figure}
 \centering
 \includegraphics[width=0.95\columnwidth]{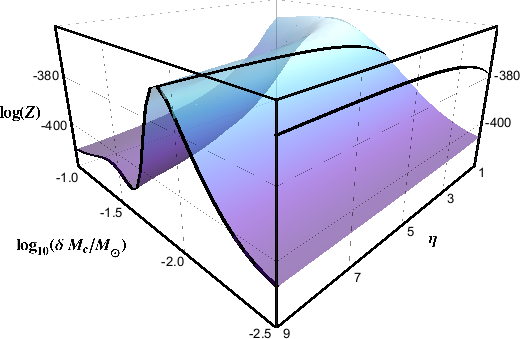}
 \caption{The hyperlikelihood, from Eq.~\eqref{eq:evtrainset}, surface for the training set ${\mathcal{D}}_{0}$ using the Mat\'ern covariance, maximised over the hyperparameter $\sigma_{f}$, plotted against the chirp mass length scale $\delta \mathcal{M}_\mathrm{c}$ and the hyperparameter, $\eta$. The hyperlikelihood does not show a clear peak, instead a ridge in the hyperparameter space favours the limiting case $\eta\rightarrow\infty$, in which limit the Mat\'ern covariance function is equal to the SE covariance function. On the near-side faces of the plot box we show the hyperlikelihood sliced parallel to the coordinate axes though the point $(\delta \mathcal{M}_\mathrm{c} = 10^{-1.9}\Msun,\eta=10)$. The solid black line on the near, left-hand face of the box very closely matches the solid black curve in Fig.~\ref{fig:Evidence_L} (up to an arbitrary additive constant).}
 \label{fig:Evidence_MAT_2Dplot}
\end{figure}

In order to explore the effect that the choice of covariance function has on the marginalised likelihood, the training sets were also interpolated using the Mat\'ern covariance function in Eq.~\eqref{eq:covMatern}. This covariance function has an additional hyperparameter, $\vec{\theta}=\{ \sigma_{f},g_{\mathcal{M}_\mathrm{c}\mathcal{M}_\mathrm{c}}, \eta \}$. The hyperlikelihood for training set ${\mathcal{D}}_{0}$ was maximised for this covariance function. It was found that the hyperlikelihood surface did not possess a peak, instead a ridge was found tending to a maximum at a value $\eta\rightarrow\infty$, and values of $\sigma_{f}$ and $g_{\mathcal{M}_\mathrm{c}\mathcal{M}_\mathrm{c}}$ were found to be the same as for the SE covariance function. In Fig.~\ref{fig:Evidence_MAT_2Dplot} we plot the log-hyperlikelihood (maximised over $\sigma_{f}$) against chirp-mass length scale and the additional hyperparameter $\eta$.

As the Mat\'ern covariance function recovers the SE function in the limit $\eta\rightarrow\infty$, there will be no difference in the performance of the interpolants for this training set when using the Mat\'ern or SE covariance functions. If the volume under the hyperlikelihood surface (the hyperevidence) is used as a figure-of-merit for which covariance function the data favours, then in this case the data is equally well described by either covariance function, but the SE covariance function is favoured over the Mat\'ern due to the smaller prior volume (the Occam penalty).

The hyperlikelihood was also calculated for both training sets ${\mathcal{D}}_{0}$ and ${\mathcal{D}}_{1}$ using the PLE covariance, see Eq.~\eqref{eq:covPL}, and the Cauchy covariance, see Eq.~\eqref{eq:covCauchy}, considered in Sec.~\ref{sec:cov}. In both cases a similar behaviour was observed. For the PLE covariance, a peak in the hyperlikelihood was found at $\eta =2$, where the PLE covariance equals the SE covariance. For the Cauchy covariance, a ridge in the hyperlikelihood was found tending to a maximum for $\eta\rightarrow\infty$ (similar to the Mat\'ern case shown in Fig.~\ref{fig:Evidence_MAT_2Dplot}), in which limit the Cauchy covariance also recovers the SE covariance. As with the Mat\'ern covariance, if the hyperlikelihood is used as a figure-of-merit for selecting the covariance function then the SE covariance is favoured over both the PLE and Cauchy functions due to the Occam penalty.

It is clear that interpolations of the training sets ${\mathcal{D}}_{0}$ and ${\mathcal{D}}_{1}$ using any of the PLE, Cauchy, or Mat\'ern covariance functions, evaluated at the hyperlikelihood-maximising hyperparameters, would yield identical results to an interpolation using the simpler SE covariance. For this reason, in the following sections we do not use the PLE, Cauchy, or Mat\'ern functions further and instead focus on the SE covariance function. We will, however, also consider using the Wendland polynomial function in the following sections as it reduces the computational cost.

The hyperlikelihood for the compact support Wendland polynomial covariance functions are shown in Fig.~\ref{fig:Evidence_MAT_2Dplot}, for the cases $q=0,\,1,\,2,\,3$. The compact-support functions can develop multiple-peaks in the hyperlikelihood surface associated with the length-scale of the training set: multiples of the training-set grid spacing are indicated with vertical blue lines in Fig.~\ref{fig:Evidence_MAT_2Dplot}. These subsidiary peaks occur in the $\delta \mathcal{M}_\mathrm{c}$ hyperparameter because as the size of the compact-support region grows, the (integer) number of training set points it contains changes discontinuously.

\begin{figure}
 \centering
 \includegraphics[width=0.5\textwidth]{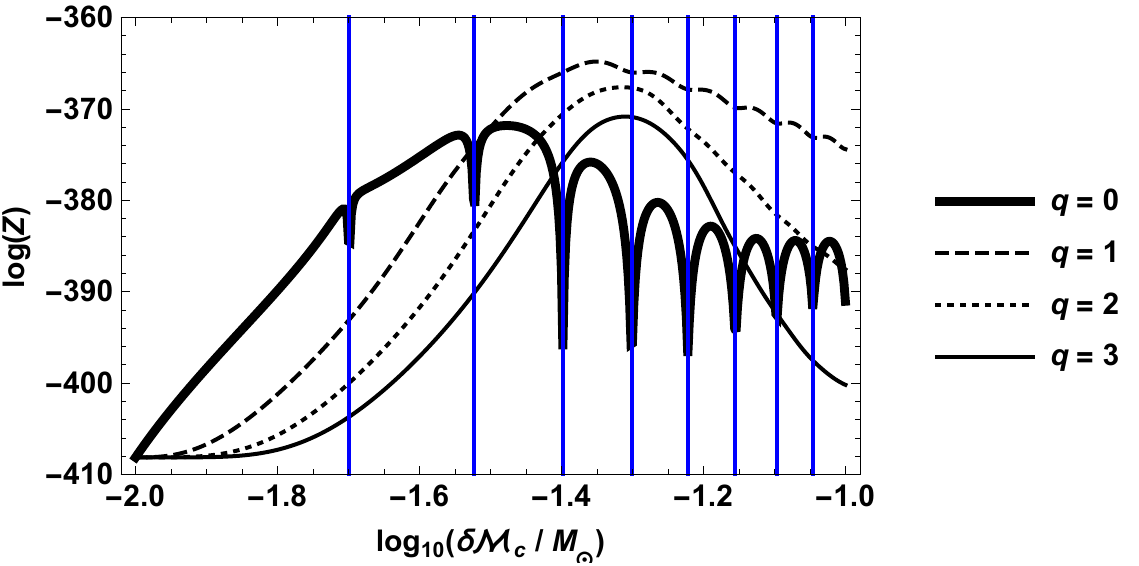}
 \caption{The hyperlikelihood, from Eq.~\eqref{eq:evtrainset}, for the training set ${\mathcal{D}}_{0}$ using the Wendland polynomial covariance functions, maximised over the scale hyperparameter $\sigma_{f}$, plotted against the chirp-mass length scale $\delta \mathcal{M}_\mathrm{c}$. The vertical blue lines indicate multiples of the training-set grid spacing $\Delta \mathcal{M}_\mathrm{c}$.}
 \label{fig:Evidence_ALLq}
\end{figure}

From Fig.~\ref{fig:Evidence_ALLq} it can be seen that for the training set ${\mathcal{D}}_{0}$, a value of $q=1$ is favoured with a length-scale $\delta \mathcal{M}_\mathrm{c}=4.37\times 10^{-2}\Msun$. In the following sections we will use Wendland covariance function with all values of $q$ (and their associated peak hyperlikelihood length scales) to interpolate ${\mathcal{D}}_{0}$. 

The optimum hyperparameters depend on the detector noise power spectral density via the overlap matrix $\langle \delta h(\vec{\lambda}_{i})|\delta h(\vec{\lambda}_{j})\rangle$. In App.~\ref{sec:diffPSD}, an investigation of the sensitivity of the optimum hyperparameters to small changes in the detector noise properties is described. It was found that for any realistic changes to the noise curve, the optimum hyperparameters were changed by an amount too small to have any noticeable effect on the interpolant.

\subsection{The interpolated waveforms}\label{sec:wavematch}

The GPR waveform $H(\vec{\lambda})-\mu(\vec{\lambda})$ could be viewed as a new waveform approximant formed from the approximant waveforms and the use of GPR on the training set of accurate waveforms. It is then natural to ask how this new approximant compares to the original ones. This can be assessed by calculating the overlap between the different waveforms, where the overlap is defined by
\begin{equation}
\mathrm{overlap}(a,b)=\frac{\left\langle  a\middle|b\right\rangle}{\left\| a\right\|\left\| b\right\|}\,,
\end{equation}
using the inner product defined in Eq.~\eqref{eq:innerprod}. 

\begin{figure}
 \centering
 \includegraphics[width=0.5\textwidth]{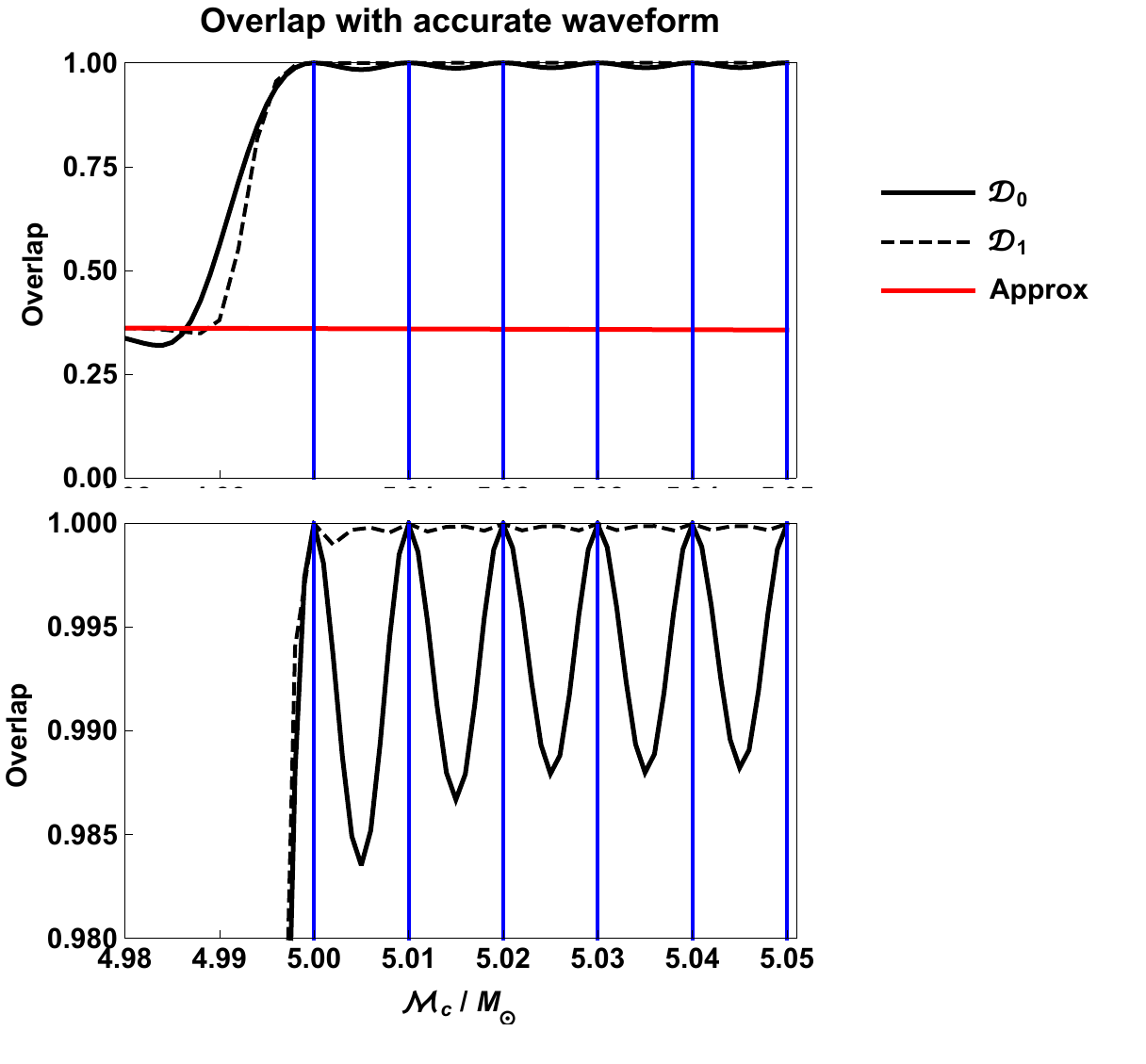}
 \caption{A plot of the overlap between the interpolated waveform $H(\vec{\lambda})-\mu(\vec{\lambda})$ and the accurate waveform $h(\vec{\lambda})$ as a function of the chirp mass $\mathcal{M}_\mathrm{c}$. The bottom panel is the same plot with a different ordinate axis scale. The two black lines show the overlap using both training sets, ${\mathcal{D}}_{0}$ and ${\mathcal{D}}_{1}$, interpolated using the SE covariance function. The red line shows the overlap between the approximate waveform $H(\vec{\lambda})$ and the accurate waveform $h(\vec{\lambda})$ for comparison. The vertical blue lines show the position of the training set points for ${\mathcal{D}}_{0}$. In the bottom panel, it can be seen that, for either interpolant, the overlap becomes one when evaluated at the training set points.} 
\label{fig:match_sets}
\end{figure}

\begin{figure}
 \centering
 \includegraphics[width=0.5\textwidth]{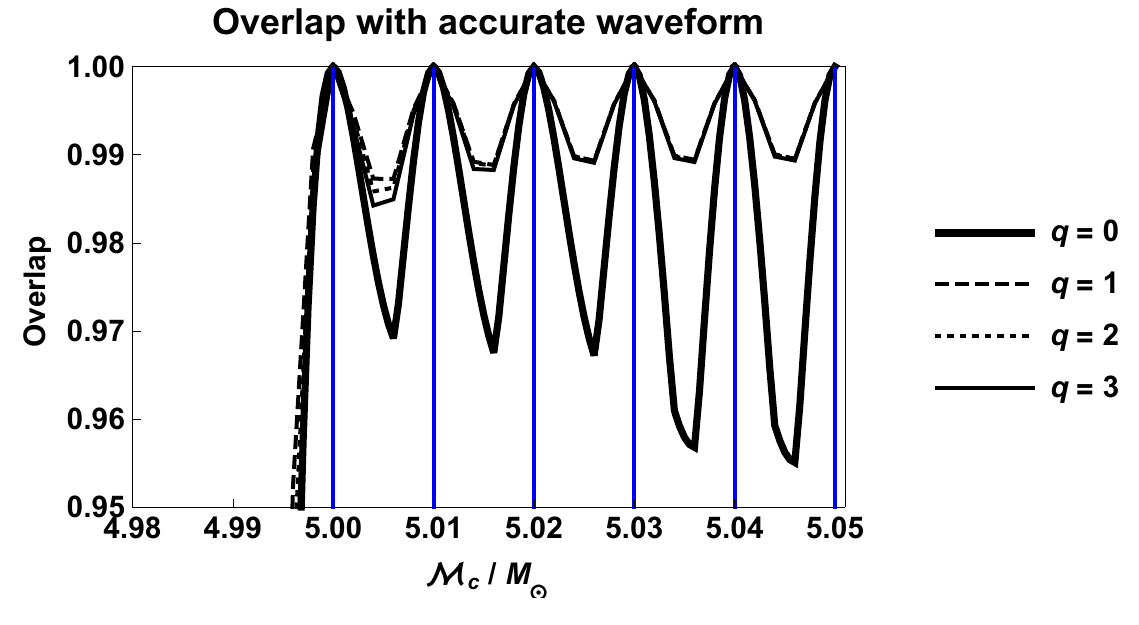}
 \caption{A plot of the overlap (or overlap) between the interpolated waveform $H(\vec{\lambda})-\mu(\vec{\lambda})$ and the accurate waveform $h(\vec{\lambda})$ as a function of the chirp mass $\mathcal{M}_\mathrm{c}$. The different curves correspond to using the Wendland polynomial covariance functions with different values of $q$ to interpolate the training set ${\mathcal{D}}_{0}$. The vertical blue lines show the position of the training set points for ${\mathcal{D}}_{0}$.} 
\label{fig:WENDLANDmatch}
\end{figure}

Only considering the overlap misses the important extra benefit which the marginalised likelihood approach brings. Our method is not just supplying a new waveform approximant, but also providing a way of modifying the posterior to account for the uncertainties known to be in the approximant. This extra information which modifies the likelihood surface is included through $\sigma(\vec{\lambda})$. Nonetheless, it is still informative to temporarily treat $H(\vec{\lambda})-\mu(\vec{\lambda})$ as if it were a new waveform approximant and see how it compares with the approximants $h(\vec{\lambda})$ and $H(\vec{\lambda})$ from which it was built. Fig.~\ref{fig:match_sets} shows the waveform overlap between the interpolated waveform $H(\vec{\lambda})-\mu(\vec{\lambda})$ and the accurate waveform $h(\vec{\lambda})$ as a function of chirp mass near the edge of the training set. Also shown in the dotted curve is the overlap between the approximate waveform $H(\vec{\lambda})$ and the accurate waveform $h(\vec{\lambda})$. The interpolated waveforms have a much higher overlap than the approximate waveforms, as would be expected. Within the training set the overlap is increased from $\sim 0.35$ to no less than $\sim 0.985$ even for the sparser training set $D_{0}$. For the denser training set ${\mathcal{D}}_{1}$ overlaps no worse than $\sim 0.999$ were found inside the range of the training set. Outside the training set the interpolated waveform tends rapidly to the approximate waveform $H(\vec{\lambda})$. 

The training set waveforms were also interpolated using the Wendland compact support covariance functions discussed in Sec.~\ref{subsec:WENDLANDcovfuncs}. The cases $q=0,\,1,\,2,\,3$ were considered separately. The waveform overlap using these interpolants is plotted in Fig.~\ref{fig:WENDLANDmatch}. The performance of these interpolants should be compared with the results using the SE covariance function in Fig.~\ref{fig:match_sets}.

The least smooth of the Wendland polynomials, the $q=0$ case, performs noticeably worse than the SE covariance; inside the training set the overlap drops as low as $\sim 0.955$ compared to $\sim 0.985$ for the SE. However, even a overlap of $\sim 0.955$ is still a great improvement over the overlap of $\sim 0.35$ for the approximate waveform alone. For the $q=0$ Wendland polynomial the interpolant has a discontinuous first derivative, which can be seen in Fig.~\ref{fig:WENDLANDmatch} (this is expected and was discussed in Sec.~\ref{sec:cov} and in detail in App.~\ref{app:A}). The higher values of $q$ have discontinuities in the higher ordered derivatives, but these curves look smooth to the eye. The smoother Wendland polynomials, with $q>0$, all perform very similarly to the SE covariance function; inside the training set the overlap drops as low as $\sim 0.985$ for the $q=2$ interpolant.

\subsection{The GPR uncertainty}\label{sec:waveerr}

\begin{figure}
 \centering
 \includegraphics[width=0.45\textwidth]{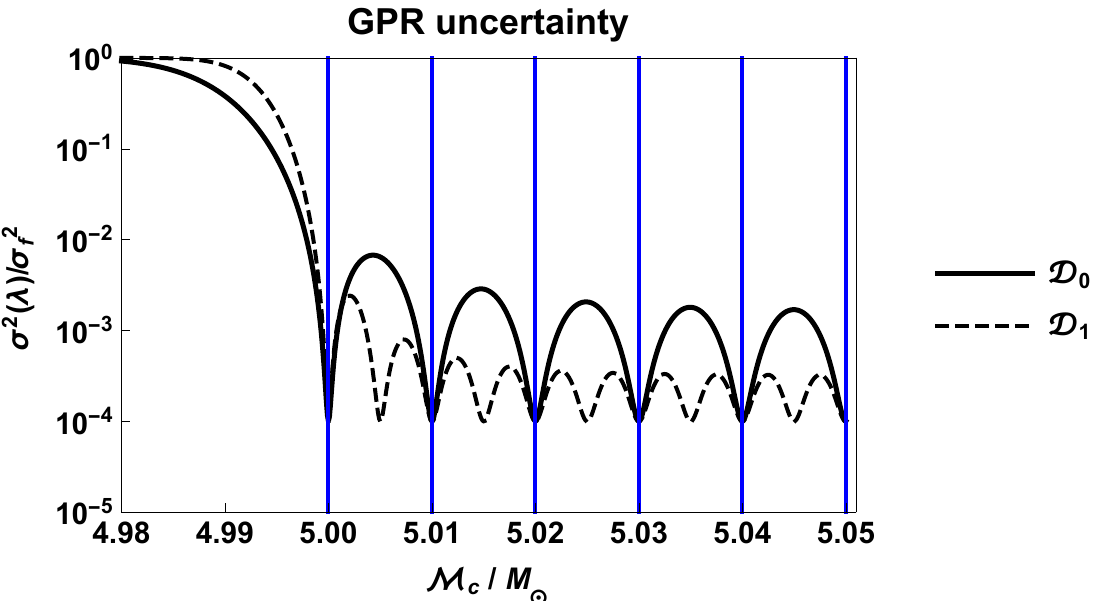}
 \caption{A plot of the GPR uncertainty $\sigma^{2}(\vec{\lambda})$ as a function of the chirp mass parameter for both of the training sets, using the SE covariance function. The vertical blue lines show the position of the training set points for ${\mathcal{D}}_{0}$. Outside of the training set the uncertainty tends to a constant $\simga_{f}^{2}$. Inside the training sets the error is approximately periodic with minima at the training set points. The maximum uncertainty inside the training set is smaller for the denser training sets. }
 \label{fig:sigma_sets}
\end{figure}

The GPR performs an interpolation of the points in the training set and naturally returns a Gaussian error $\sigma(\vec{\lambda})$, see Eq.~\eqref{eq:GPRsigma2}, for each interpolated point. In our present one-dimensional interpolation this is simply a function of $\mathcal{M}_\mathrm{c}$. A small section of this curve taken from the edge of the training set is shown in Fig.~\ref{fig:sigma_sets}. Inside the training set, the error surface has a regular, periodic pattern with minima at the training set points and maxima in between. This regularity is because the GP used for the interpolation is stationary, the training-set points used are regularly spaced, and each point has an identical error (a jitter $J=10^{-4}$). If these conditions were to be relaxed, then the error surface would become more complicated. In general, a larger $\sigma(\vec{\lambda})$ indicates greater theoretical uncertainty and highlights regions where we would benefit from additional accurate waveforms (e.g., where it would be beneficial to perform more NR simulations).

Near the edge of the training set the behaviour becomes less regular and well outside of the training set the error tends to a constant value, $\sigma^{2}(\vec{\lambda})\rightarrow\sigma_{f}$ as $\vec{\lambda}\rightarrow\infty$. This behaviour is seen in Fig.~\ref{fig:sigma_sets} for all three training sets. The training sets with smaller grid spacings have smaller uncertainties everywhere in parameter space. 

\begin{figure}
 \centering
 \includegraphics[width=0.45\textwidth]{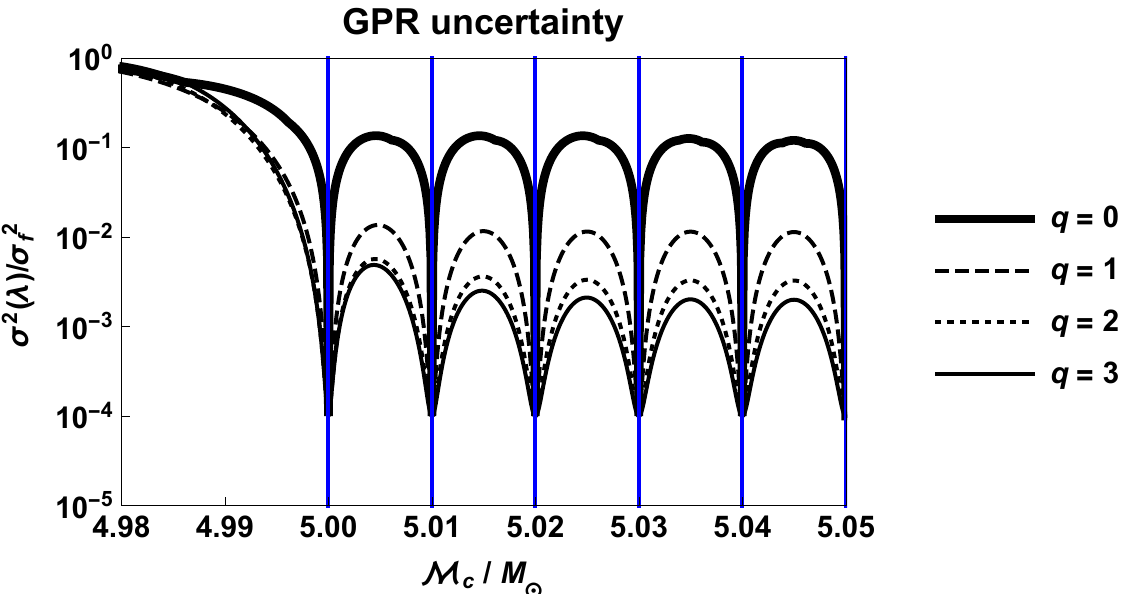}
 \caption{A plot of the GPR uncertainty $\sigma^{2}(\vec{\lambda})$ as a function of the chirp mass parameter for the training set ${\mathcal{D}}_{0}$, using the Wendland polynomial covariance functions. The vertical blue lines show the position of the training set points. }
 \label{fig:WENDLANDsigma}
\end{figure}

The GPR uncertainty was also calculated using the Wendland polynomial covariance functions to interpolate the training set ${\mathcal{D}}_{0}$; these are shown in Fig.~\ref{fig:WENDLANDsigma}. The GPR uncertainty, expressed as a fraction of $\sigma_{f}^{2}$, is largest for the smallest values of $q$; this can be traced back to the optimum length scale for the Wendland polynomials increasing with $q$ (see Fig.~\ref{fig:Evidence_ALLq}). This means that the uncertainty grows more slowly as the interpolating point moves away from the training set points, and hence reaches a smaller maximum value between training set points. The smoother ($q>0$) Wendland polynomials perform similarly to the SE covariance function, in the sense that both the GPR interpolants (which we quantify via the overlap) and the GPR uncertainties are almost identical. Hence, in the following sections we will only consider using the SE covariance function; the high $q$ Wendland polynomials would yield identical results.

\subsection{The likelihood}\label{sec:likelihoodresults}

Finally we put together the interpolated waveform $H(\vec{\lambda})-\mu(\vec{\lambda})$ and the GPR uncertainty $\sigma^{2}(\vec{\lambda})$ to give the marginalised likelihood in Eq.~\eqref{eq:finalresult}. We compare the performance of the marginalised likelihood ${\mathcal{L}}(\vec{\lambda})$ to the approximate likelihood $L(\vec{\lambda})$ and the accurate likelihood $L'(\vec{\lambda})$. For the injected signal we use the accurate waveform $h(\vec{\lambda})$. We also consider the case where the noise realisation is zero (the most likely realisation), this makes comparisons easier. 

\begin{figure*}
 \centering
 \includegraphics[trim={1cm 0.3cm 0cm 0.3cm},width=1.0\textwidth]{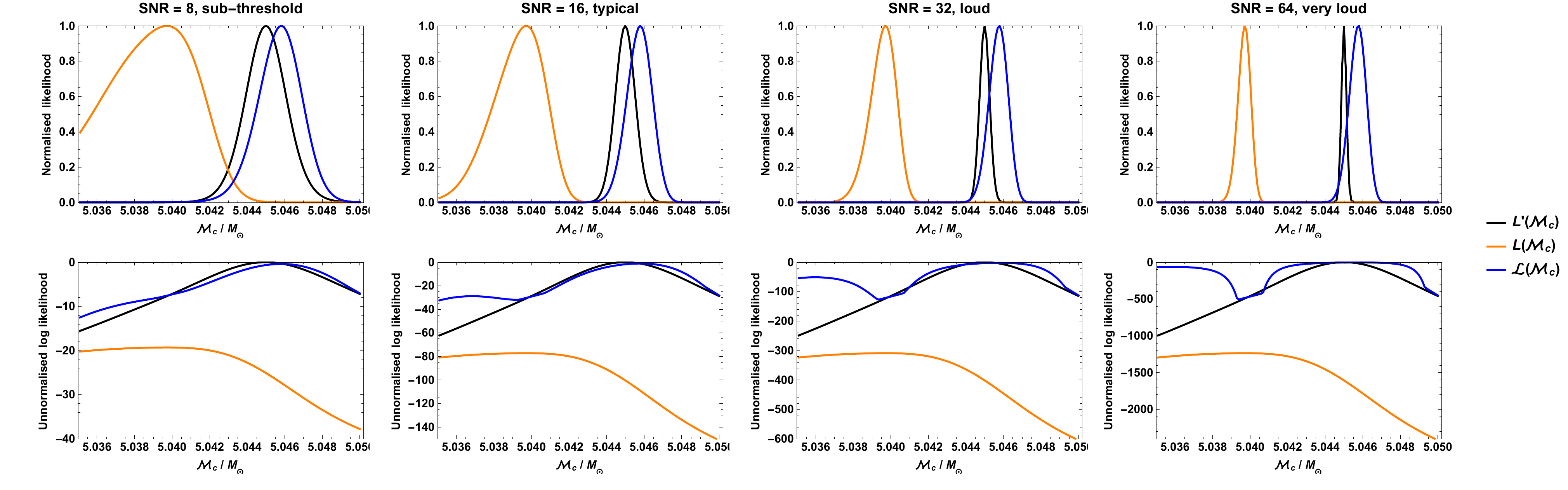}
 \caption{A plot of the different likelihoods for a variety of SNRs. Vertical lines indicate the position of training set points. The top row of panels show the likelihood normalised to the same peak value; this makes the peak positions clear and shows how the marginalised likelihood tackles the parameter estimation problem associated with the inaccurate models. The bottom row of panels shows the log-likelihood; this makes the suppression of the peak value of the approximate likelihood clear and shows how the marginalised likelihood could be used to tackle the detection problem. } 
 \label{fig:DifferentSNR}
\end{figure*}

We injected a signal at a chirp mass of $\mathcal{M}_\mathrm{c}=5.045 \Msun$; this is inside the training set ${\mathcal{D}}_{0}$ and midway between training set points. Injecting the signal midway between the points is conservative as this is the point at which we would expect the marginalised likelihood to perform worst. The three different likelihoods were evaluated as a function of chirp mass (all other parameters set to the injected values). This was done at a range of SNRs and the results are shown in Fig.~\ref{fig:DifferentSNR}. The top row of panels in Fig.~\ref{fig:DifferentSNR} show the likelihoods renormalised to a peak value of one, this makes the relative positions of the peaks clear and easy to compare. The bottom row of panels shows the log-likelihood without any renormalisation, this illustrates how the approximate likelihood is suppressed relative to the true likelihood (the detection problem discussed in Sec.~\ref{intro}).

The exact likelihood $L'(\vec{\lambda})$ is always peaked at the injected value of the chirp mass (because the injected noise realisation is zero) and the width of the peak decreases with increasing SNR. The approximate likelihood $L(\vec{\lambda})$ is peaked way from the true value, indicating a systematic error of $\Delta_{\mathrm{sys}}\mathcal{M}_\mathrm{c}=5.2\times 10^{-3}\Msun$. The width of the approximate likelihood peak also decreases with increasing SNR and for SNR $\gtrsim 12$ (which is also roughly the detection threshold \cite{2013arXiv1304.0670L,2015ApJ...804..114B}) the true parameters are excluded at increasing significance. The bottom row of panels in Fig.~\ref{fig:DifferentSNR} shows that the approximate likelihood is suppressed by a significant amount, for a typical SNR of 16 it is supressed by $80$ in log relative to the exact likelihood; this reduces the Bayesian evidence for a detection. The factor by which the approximate likelihood is suppressed increases exponentially with SNR. Finally, the marginalised likelihood is peaked much closer to the exact likelihood: the systematic error is reduced to $\Delta_{\mathrm{sys}}\mathcal{M}_\mathrm{c}=9.0\times 10^{-4}\Msun$. However, as discussed in Sec.~\ref{sec:infSNR}, the peak in the marginalised likelihood does not continually narrow as the SNR increases; for SNR $\gtrsim30$ the width becomes constant. Consequently, the true parameters are never excluded at high significance; in the limit of infinite SNR the true parameters lie at the $\sim 1\sigma$ level. The bottom panel of Fig.~\ref{fig:DifferentSNR} shows that the marginalised likelihood is not suppressed relative to the exact likelihood in the vicinity of the peak.

Comparing the properly normalised likelihoods, we see that the marginalised and exact likelihoods roughly agree at low SNR. As the SNR is increased, the marginalised likelihood deviates from the exact likelihood and develops oscillatory behaviour with period equal to the training set point spacing. In the limit of low SNR, all of the parameter estimation uncertainty comes from the noise, but as the SNR increases, the relative size of this statistical uncertainty becomes smaller and at high SNR we are dominated by model uncertainty. The marginalised likelihood correctly encapsulates this behaviour, as can be seen in the sequence from left to right in Fig.~\ref{fig:DifferentSNR}.

\section{Summary}\label{sec:discussion}

In \cite{Letter}, some of the authors suggested GPR as a means of incorporating theoretical uncertainty into GW data analysis. We have now thoroughly investigated the properties of this method, elucidating considerations for a practical implementation. A detailed derivation of the marginalised likelihood, and the use of GPR to interpolate model error was presented in Sec.~\ref{sec:method}. GPR is non-parametric, in the sense that only the functional form of the covariance function is specified by hand, with its hyperparameters then learnt from the training set, making it well suited to modelling theoretical uncertainty. The expression for the marginalised likelihood derived in Sec.~\ref{sec:method} made some assumptions about the frequency covariance of the waveforms in the training set; in particular it was assumed that at a particular point in parameter space, the model error is highly correlated in frequency. These assumptions may prove to be too restrictive in the future, and could be relaxed by simultaneously performing GPR interpolation in frequency (as well as in parameter space) on the training set waveforms; this would have the added advantage of allowing the inclusion of waveforms with different frequency samplings, which may be beneficial when using multiple waveform approximants and NR waveforms from multiple sources.

The choice of covariance function is central to GPR as it encodes our prior beliefs about the function space that we are interpolating. We discussed various choices of covariance function in Sec.~\ref{sec:cov}. We have found that the simple SE covariance function (as used in \cite{Letter}) performs as well as more complicated alternatives, at least for the relatively small one dimensional training sets considered here. The compact-support Wendland covariance functions with large $q$ were found to perform comparably to the SE, but offer the additional advantage of reduced computational cost. This makes them appealing for future work involving larger training sets.

We proved a number of properties for our marginalised likelihood in Sec.~\ref{sec:props}, in particular its limiting behaviour for large signal amplitude (where the theoretical errors are known to be most significant \cite{PhysRevD.76.104018}) and its limiting behaviour both far from and near a point in the training set. In the discussion of the latter, the linearised results previously obtained in \cite{PhysRevD.76.104018} were recovered. All of these properties demonstrate the suitability of GPR for making robust inferences. The marginalised likelihood successfully describes our belief in our inferences, including our uncertainty in waveform templates.

In Sec.~\ref{sec:allres}, we presented a one-dimensional implementation of our marginalised likelihood and demonstrated that it offers an improvement in PE accuracy. For this, we chose two inexpensive waveforms to aid computation; in real data analysis situations, we expect more accurate waveforms (including those calculated using NR) to be used in the training. However, this choice of waveforms does illustrate the efficacy of the new marginalised likelihood. In particular we find that even when using qualitatively different waveforms (inspiral-only TaylorF2 compared to inspiral--merger--ringdown IMRPhenomC), waveform matches as high as $\sim98.5\%$ can be obtained in the mass range we considered. Restricting ourselves to the simple case of one-dimensional PE, we explored various possibilities for GPR. In particular, the effect of different training set sizes was  examined; as expected, the performance of the marginalised likelihood is improved by using denser training sets. Additionally, the impact of varying the SNR of the injected waveform was studied. In the standard likelihood model, errors become more severe as the SNR is increased, but we confirmed that even in the limit of large SNR, the marginalised likelihood remained consistent with the injected parameters. We expect these results to carry over to a full multidimensional analysis, which is the next step in developing this technique.

A possibly complementary use for the GPR approximant is as a less expensive alternative to the accurate waveform for more expedient PE; however, there exist other means of constructing computationally inexpensive waveforms, such as reduced-order modelling~\cite{2015PhRvL.114g1104C,2014CQGra..31s5010P}. The advantage of GPR is that it not only supplies an interpolant, but also gives an uncertainty, which can be used to gauge accuracy away from training points. A second, related feature is that GPR naturally allows for uncertainty to be included in the accurate models that the interpolant is calibrated against. It is the ability of GPR to include theoretical uncertainty that makes it attractive for GW astronomy, that this can be done without significant on-line cost is a welcome bonus.

In conclusion, marginalising over waveform uncertainty is a robust and effective method of accounting for theoretical error in both PE and detection problems. GPR is a natural and effective means of performing this marginalisation. The marginalised likelihood is naturally inferior to a likelihood calculated with more accurate (but inevitably more computationally expensive) waveforms, but it offers significantly improved performance over the standard likelihood calculated with cheap waveforms. In addition, the marginalised likelihood is almost as quick to evaluate online as the standard likelihood, although there is additional offline computation required to construct the training set and train the Gaussian process.

\acknowledgments{CJM and CPLB are supported by the STFC. AJKC's work is supported by the Cambridge Commonwealth, European and International Trust. JRG's work is supported by the Royal Society. We thank Will Farr, Alberto Vecchio, Ilya Mandel, Ben Farr, Daniel Holz and the Compact Binary Coalescence PE Group for useful discussions. This work made use of computational resources provided by the Leonard E.\ Parker Center for Gravitation, Cosmology and Astrophysics at University of Wisconsin--Milwaukee. LIGO was constructed by the California Institute of Technology and Massachusetts Institute of Technology with funding from the National Science Foundation and operates under cooperative agreement PHY-0757058. This document has been assigned LIGO document reference LIGO-P1500162.}

\bibliography{bibliography}

\appendix
\section{Continuity and differentiability of GPs}\label{app:A}
In this appendix we give proofs of the results stated in Sec.~\ref{sec:cov} concerning the continuity and differentiability of GPs, following the approach of \cite{Adler}.
Let $\vec{\lambda}_{1},\vec{\lambda}_{2},\vec{\lambda}_{3}\ldots$ be a sequence of points in parameter space which converges to a point $\vec{\lambda}_{*}$, in the sense $\lim_{\ell\rightarrow\infty}~|\vec{\lambda}_{\ell}~-~\vec{\lambda}_{*}| = 0$, where, as in Section~\ref{sec:cov}, $|\vec{x}|$ denotes the norm with respect to the metric on parameter space, as discussed in Section~\ref{subsec:metric}. The GP $Y (\vec{\lambda})$ is said to be MS continuous at $\vec{\lambda}_{*}$ if
\begin{equation}\label{eq:limMS}
\lim_{\ell\rightarrow\infty}\Ex\left[\left( Y(\vec{\lambda}_{\ell})-Y(\vec{\lambda}_{*})\middle|Y(\vec{\lambda}_{\ell})-Y(\vec{\lambda}_{*})\right)\right]=0 \,,
\end{equation}
where $\Ex[\ldots]$ denotes the expectation of the enclosed quantity over realisations of the GP. For notational convenience, we denote this MS limit as
\begin{equation}
Y(\vec{\lambda}_{*}) =  \underset{\ell \rightarrow \infty}{\mathrm{l.i.m.}}\,Y(\vec{\lambda}_{\ell})\,,
\end{equation}
where $\mathrm{l.i.m.}$ stands for limit in mean \cite{Wiener1949}. MS continuity implies continuity in the mean,
\begin{equation} \label{eq:limM}
\lim_{\ell\rightarrow\infty}\Ex\left[Y(\vec{\lambda}_{\ell})-Y(\vec{\lambda}_{*})\right]=0\,.
\end{equation}
This follows from considering the variance of the quantity $Y(\vec{\lambda}_{\ell})-Y (\vec{\lambda}_{*})$, and the fact that variance is non-negative. There are other notions of continuity of GPs used in the literature, but the notion of MS continuity relates most easily to the covariance. The mean and the covariance of a GP are defined as
\begin{eqnarray}
m(\vec{\lambda}) &=&\Ex\left[Y (\vec{\lambda})\right]\,, \\
k(\vec{\lambda}_{1},\vec{\lambda}_{2})&=&{\Ex\left[\left(  Y(\vec{\lambda}_{1})-m(\vec{\lambda}_{1})\middle|Y(\vec{\lambda}_{2})-m(\vec{\lambda}_{2})\right)\right]}\,.\nonumber
\end{eqnarray}
Using these, Eq.~\eqref{eq:limMS} can be written as
\begin{eqnarray}
\lim_{\ell\rightarrow\infty}\left\{k(\vec{\lambda}_{*},\vec{\lambda}_{*})-2k(\vec{\lambda}_{\ell},\vec{\lambda}_{*}) + k(\vec{\lambda}_{\ell},\vec{\lambda}_{\ell})\right. \: && \nonumber\\*
+\left.\left( m(\vec{\lambda}_{*})-m(\vec{\lambda}_{\ell})\middle|m(\vec{\lambda}_{*})-m(\vec{\lambda}_{\ell})\right)\right\} &=&0 \,,
\end{eqnarray}
and using the continuity of the mean in Eq.~\eqref{eq:limM} gives
\begin{equation}
\label{eq:continuityofcov}
\lim_{\ell\rightarrow\infty}\left[k(\vec{\lambda}_{*},\vec{\lambda}_{*})-2k(\vec{\lambda}_{\ell},\vec{\lambda}_{*})+k(\vec{\lambda}_{\ell},\vec{\lambda}_{\ell})\right]=0\,.
\end{equation}
This condition is satisfied if the covariance function is continuous at the point $\vec{\lambda}_{1}=\vec{\lambda}_{2}=\vec{\lambda}_{*}$. Therefore, we arrive at the result that if the covariance function is continuous in the usual sense at some point $\vec{\lambda}_{*}$, then the corresponding GP is MS continuous at this point. In fact, a GP is continuous in MS if \emph{and only if} the covariance function is continuous \cite{Adler}, although this is not proved here. In the special case of stationary covariance this reduces to checking continuity of $k(\vec{\tau})$ at $\vec{\tau}=0$, and in the special case of isotropic covariance, continuity of $k(\tau)$ at $\tau=0$.

We now move on from continuity to consider differentiability. In the spirit of Eq.~\eqref{eq:limMS}, the notion of taking the MS derivative of a GP is defined as
\begin{eqnarray}
\frac{\partial Y(\vec{\lambda})}{\partial \vec{\lambda}^{a}} &=& \underset{\epsilon \rightarrow 0}{\mathrm{l.i.m.}}\,X_{a}(\vec{\lambda},\epsilon)\,, \\
\textrm{where }X_{a}(\vec{\lambda},\epsilon) &=& \frac{Y(\vec{\lambda}+\epsilon \,\hat{e}_{a})-Y(\vec{\lambda})}{\epsilon}\,\label{eq:derivGPdef}
\end{eqnarray}
with parameter-space unit vector $\hat{e}_{a}$. This definition can be easily extended to higher-order derivatives \cite{Adler}.
\begin{figure*}
 \centering
 \includegraphics[width=0.95\textwidth]{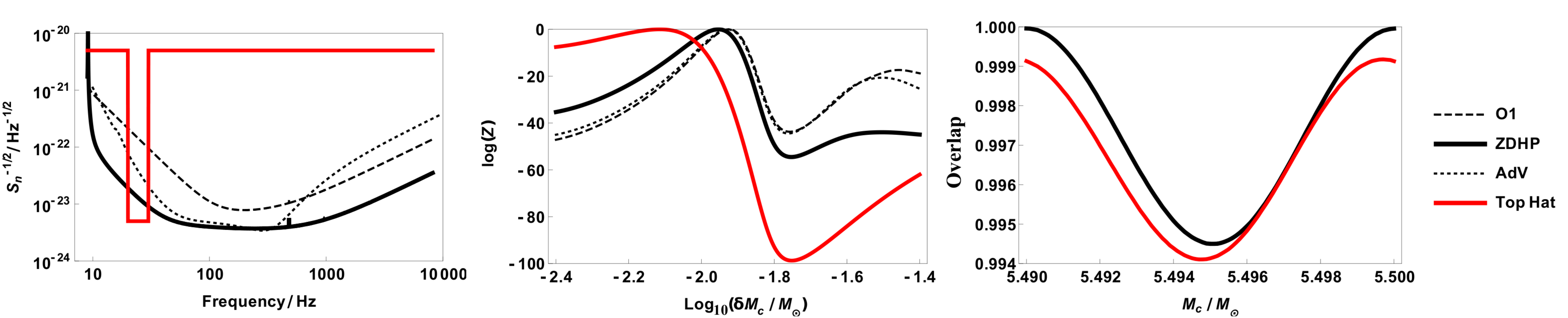}
 \caption{The left-hand panel shows three different noise curves for ground-based detectors in the advanced era. Also shown in red is an unrealistic noise which we use for comparison. The centre-panel shows the hyper-likelihood surface for the training set ${\mathcal{D}}_{0}$ using the SE covariance function and with the overlap matrix calculated using each of the noise curves in the left-hand panel. The right-hand panel shows the waveform overlap between the accurate and the interpolated waveform evaluated for parameter values between two training set points. The interpolants based on the realistic noise curves perform equally well (the curves lie on top of each other). The unrealistic noise curve performs worst, but still gives overlaps greater than $0.994$.}
 \label{fig:DIFF}
\end{figure*}
The MS derivative of a GP is also a GP; this follows simply from the fact that the sum of Gaussians is also distributed as a Gaussian. The covariance of $X_{a}(\vec{\lambda},\epsilon)$ is given by
\begin{eqnarray}
K_{\epsilon}(\vec{\lambda}_{1},\vec{\lambda}_{2})&=&\Ex\left[\left( X_{a}(\vec{\lambda}_{1},\epsilon)-\Xi(\vec{\lambda}_{1},\epsilon)\middle|\right.\right.\nonumber\\*
&&\left.\left.\quad\quad X_{a}(\vec{\lambda}_{2},\epsilon)-\Xi(\vec{\lambda}_{2},\epsilon)\right)\right] \,\,
\end{eqnarray}
where $\Xi_{a}(\vec{\lambda},\epsilon)=\Ex[X_{a}(\vec{\lambda},\epsilon)]$. It then follows that
\begin{eqnarray}
K_{\epsilon}(\vec{\lambda}_{1},\vec{\lambda}_{2})&=&\frac{k(\vec{\lambda}_{1}+\epsilon,\vec{\lambda}_{2}+\epsilon)-k(\vec{\lambda}_{1},\vec{\lambda}_{2}+\epsilon)}{\epsilon^{2}}\nonumber\\
&&-\frac{k(\vec{\lambda}_{1}+\epsilon,\vec{\lambda}_{2})-k(\vec{\lambda}_{1},\vec{\lambda}_{2})}{\epsilon^{2}} \,.\label{eq:covderivs}
\end{eqnarray}
Substituting this into Eq.~\eqref{eq:derivGPdef}, the limit in MS becomes a normal limit, and the result is obtained that the MS derivative of a MS continuous GP with covariance function $k(\vec{\lambda}_{1},\vec{\lambda}_{2})$ is a GP with covariance function $\partial^{2} k(\vec{\lambda}_{1},\vec{\lambda}_{2})/\partial\vec{\lambda}_{1}^{a}\partial\vec{\lambda}_{2}^{a}$. In general the covariance function of the $n_{d}$-times MS differentiated GP
\begin{equation}\label{eq:GPderiv}
\frac{\partial^{n_{d}}Y(\vec{\lambda})}{\partial\vec{\lambda}^{a_{1}}\partial\vec{\lambda}^{a_{2}}\ldots\partial\vec{\lambda}^{a_{n_{d}}}}\,,
\end{equation}
is given by the $2n_{d}$-times differentiated function
\begin{equation}\label{eq:Covderiv}
\frac{\partial^{2n_{d}}k(\vec{\lambda}_{1},\vec{\lambda}_{2})}{\partial\vec{\lambda}_{1}^{a_{1}}\partial\vec{\lambda}_{2}^{a_{1}}\partial\vec{\lambda}_{1}^{a_{2}}\partial\vec{\lambda}_{2}^{a_{2}}\ldots\partial\vec{\lambda}_{1}^{a_{n_{d}}}\partial\vec{\lambda}_{2}^{a_{n_{d}}}} \, .
\end{equation}
From the above results relating the MS continuity of GPs to the continuity of the covariance function at ${\vec{\lambda}_{1}=\vec{\lambda}_{2}=\vec{\lambda}_{*}}$, it follows that the $n_{d}$-times MS derivative of the GP is MS continuous (the GP is said to be $n_{d}$-times MS differentiable) if the $2n_{d}$-times derivative of the covariance function is continuous at ${\vec{\lambda}_{1}=\vec{\lambda}_{2}=\vec{\lambda}_{*}}$ \cite{Stein1999}. So it is the smoothness properties of the covariance function along the diagonal points that determines the differentiability of the GP. (It can also be shown that if a covariance function is continuous at all diagonal points $\vec{\lambda}_{1}=\vec{\lambda}_{2}$ then it's everywhere continuous.)

\section{The effect of small changes in the noise PSD on the GPR interpolant}\label{sec:diffPSD}

In the offline stage of the method, the GP was trained using the hyperlikelihood in Eq.~\eqref{eq:evtrainset}. The result of this process was an interpolant which enabled fast online PE. However, this splitting into offline and online stages also has a potential problem, because the training process makes use of the overlap matrix $M_{ij}=\langle \delta h(\vec{\lambda}_{i})|\delta h(\vec{\lambda}_{j})\rangle$ which, in turn, depends upon the detector noise PSD $S_{n}(f)$. The noise PSD is not constant; it changes on short timescales as the noise drifts in the instrument (e.g., \cite{2015CQGra..32k5012A}), on longer timescales it changes more dramatically as the instrument is gradually upgraded \cite{2013arXiv1304.0670L}. There are also differences between different detectors, for example between the aLIGO and AdV instruments (or even between the two aLIGO interferometers). It would be a significant drawback if the offline training stage of the process had to be repeated for every single candidate signal because of small differences in the detector PSD.

We do not expect small changes in the noise curve to have a significant effect on the resulting interpolant. First, the noise can be rescaled by an overall constant and have no effect on the position of the peak in the hyperlikelihood; this can be seen from Eq.~\eqref{eq:evtrainset}. Second, the peak in the hyperlikelihood is typically wide, and using the hyperparameters from anywhere in the vicinity of the peak still gives reasonable, if not perfect, interpolation. Accordingly, when the PSD changes, some of the difference can be absorbed by an overall scaling, which has no effect on the results, and the remaining change shifts the peak of the hyperlikelihood away from the previously optimised values, but not enough to limit their applicability. If this is the case, then GPs trained on slightly different noise PSDs perform nearly identically to each other and there is no need to retrain for the new PSD.

To assess the sensitivity of our results to changes in the noise curve, we considered three different noise curves chosen to represent the range of possibilities in the  advanced-detector era. These are: an estimate of the observing run 1 (O1) aLIGO sensitivity (the early curve of \cite{Barsotti:2012}); the zero-detuned high-power (ZDHP) design sensitivity of aLIGO \cite{PSD:aLIGO,2015CQGra..32g4001T}, and the design sensitivity of AdV \cite{Acernese2009,Accadia2012}. As an additional check, we also considered an inverted top-hat noise curve. All of these noise curves are plotted in the left-hand panel of Fig.~\ref{fig:DIFF}. We then took the training set ${\mathcal{D}}_{0}$ and trained the SE GP to find the optimum hyperparameters. Shown in the centre-panel of Fig.~\ref{fig:DIFF} is the hyperlikelihood surface as a function of chirp mass length scale for the different noise curves. As expected, for the range of realistic noise curves the peak in the hyperlikelihood only shifts by a small amount. Finally we used the optimum hyperparameters from each of these hyperlikelihood surfaces to interpolate the training set and calculated the overlap to the accurate waveforms using the ZDHP noise curve; the results of this are shown in the right-hand panel of Fig.~\ref{fig:DIFF}. For the range of realistic noise curves, the overlap is equally good  (cf.\ \cite{Stein1999}). Although the inverted top-hat noise curve gives noticeably lower overlaps, even in that case the drop in the overlap is still less than $0.1\%$, which is smaller than the difference between the approximate and GPR likelihoods. 

This suggests it is safe to train a GP with a fixed noise curve (typical for the instruments considered). The resulting interpolants can be used to analyse all signals without worrying about small drifts in the noise.

\end{document}